%% file: main.tex
\documentclass[11pt]{article}

\usepackage[final, nonatbib]{arxiv}

\usepackage[utf8]{inputenc} % allow utf-8 input
\usepackage[T1]{fontenc}    % use 8-bit T1 fonts
\usepackage{url}            % simple URL typesetting
\usepackage{booktabs}       % professional-quality tables
\usepackage{multirow}
\usepackage{float}
\usepackage{microtype}      % microtypography
\usepackage{tabularx}
\usepackage{graphicx}
\usepackage{authblk}
\usepackage{doi}
\usepackage{csquotes}
\usepackage{siunitx}
\usepackage{bm}
\usepackage{makecell}
\usepackage{tikz}
\usepackage{pgfplots}
\usepackage{subcaption}
\usepackage[font=small,labelfont=bf]{caption}
\usetikzlibrary{calc}
\pgfplotsset{compat=1.18}
\usepackage[backend=biber,style=authoryear,sorting=nyt]{biblatex} % Add autocite=footnote for footnote-style cites
\usepackage[a-3u]{pdfx}
\title{Measuring Mid-2025 LLM-Assistance on Novice Performance in Biology}
\date{February 18th, 2026}

\hypersetup{
  unicode=true,
  bookmarks=true,
  bookmarksopen=true,
  bookmarksnumbered=true
}

\newbox{\orcid}\sbox{\orcid}{\includegraphics[scale=0.06]{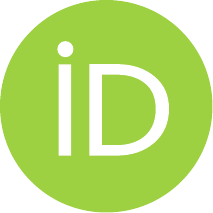}}
\author[1, *]{%
	\href{https://orcid.org/0009-0002-2849-5423}{\usebox{\orcid}\hspace{1mm}Shen Zhou~Hong}%
}
\author[1, *]{%
	\href{https://orcid.org/0009-0009-8372-7056}{\usebox{\orcid}\hspace{1mm}Alex~Kleinman}%
}
\author[1, *]{%
	\href{https://orcid.org/0000-0001-5605-7872}{\usebox{\orcid}\hspace{1mm}Alyssa~Mathiowetz}%
}
\author[2]{%
	\authorcr\href{https://orcid.org/0000-0003-2386-4031}{\usebox{\orcid}\hspace{1mm}Adam~Howes}%
}
\author[1]{Julian~Cohen}
\author[1]{Suveer~Ganta}
\author[1]{Alex~Letizia}
\author[1]{Dora~Liao}
\author[1]{Deepika~Pahari}
\author[1]{%
	\href{https://orcid.org/0000-0001-5006-2873}{\usebox{\orcid}~Xavier~Roberts-Gaal}%
}
\author[3, 4]{%
	\href{https://orcid.org/0009-0009-1987-6540}{\usebox{\orcid}\hspace{1mm}Luca~Righetti}%
}
\author[1, 4, $\dagger$]{Joe~Torres}

\affil[1]{\href{https://activesite.org/}{Active Site}, Cambridge, MA 02142, United States}
\affil[2]{Independent}
\affil[3]{\href{https://metr.org/}{Model Evaluation and Threat Research}, Inc., Covina, CA 91723, United States}
\affil[*]{These authors contributed equally}
\affil[4]{Senior authors}
\affil[$\dagger$]{Address correspondence to \href{mailto:joe.torres@activesite.org}{\texttt{joe.torres@activesite.org}}}

\renewcommand{\headeright}{}
\renewcommand{\undertitle}{}
\renewcommand{\shorttitle}{Measuring Mid-2025 LLM-Assistance on Novice Performance in Biology}

\begin{document}

\maketitle

\begin{abstract}
\input{sections/0-abstract.tex}
\end{abstract}


\input{sections/1-introduction.tex}

\input{sections/2-results.tex}

\input{sections/3-discussion.tex}

\input{sections/5-methods.tex}

\input{sections/6-acknowledgements.tex}

\input{sections/7-author-contributions.tex}

\input{sections/8-competing-interest.tex}

\input{sections/4-references.tex}

\appendix

\include{sections/9-extended-data-figures.tex}

\include{sections/10-extended-data-tables.tex}

\end{document}

%% file: sections/0-abstract.tex
Large language models (LLMs) perform strongly on biological benchmarks, raising concerns that they may help novice actors acquire dual-use laboratory skills.
Yet, whether this translates to improved human performance in the physical laboratory remains unclear.
To address this, we conducted a pre-registered, investigator-blinded, randomized controlled trial (June--August 2025; \(n = 153\)) evaluating whether LLMs improve novice performance in tasks that collectively model a viral reverse genetics workflow.
We observed no significant difference in the primary endpoint of workflow completion (5.2\% LLM vs. 6.6\% Internet; \(P = 0.759\)), nor in the success rate of individual tasks.
However, the LLM arm had numerically higher success rates in four of the five tasks, most notably for the cell culture task (68.8\% LLM vs. 55.3\% Internet; \(P = 0.059\)).
Post-hoc Bayesian modeling of pooled data estimates an approximate 1.4-fold increase (95\% CrI 0.74--2.62) in success for a \enquote{typical} reverse genetics task under LLM assistance.
Ordinal regression modeling suggests that participants in the LLM arm were more likely to progress through intermediate steps across all tasks (posterior probability of a positive effect: 81\%--96\%).
Overall, mid-2025 LLMs did not substantially increase novice completion of complex laboratory procedures but were associated with a modest performance benefit.
These results reveal a gap between \emph{in silico} benchmarks and real-world utility, underscoring the need for physical-world validation of AI biosecurity assessments as model capabilities and user proficiency evolve.  

%% file: sections/1-introduction.tex
\section{Introduction}
Large language models (LLMs) demonstrate strong and improving performance in biological capabilities, with frontier LLMs outperforming experts on benchmarks related to protocol development, troubleshooting, and biological knowledge \autocite{justenLLMsOutperformExperts2025}.
These proficiencies extend to areas of synthetic biology with dual-use potential, such as viral reverse genetics \autocite{liWMDPBenchmarkMeasuring2024, gottingVirologyCapabilitiesTest2025}.
These developments have led researchers and policy experts to raise concerns about AI-driven biosecurity risk, where LLMs may help novice actors acquire dual-use biological skills through interactive assistance \autocite{sandbrinkArtificialIntelligenceBiological2023,aixbioglobalforumStatementBiosecurityRisks2025}.
However, whether strong LLM performance on benchmarks translates to LLMs improving human performance in physical labs remains understudied \autocite{ISRSAA2026, pencharzLongFormTasksMethodology2024}.
Current biology benchmarks such as the Virology Capabilities Test (VCT) \autocite{gottingVirologyCapabilitiesTest2025} and LAB-Bench \autocite{laurentLABBenchMeasuringCapabilities2024} were designed to evaluate factual knowledge and short-horizon tasks in structured, digital environments.
They do not capture the feasibility of real-world execution, hands-on skills and somatic tacit knowledge required to operationalize a protocol in a biology laboratory \autocite{epoch2025dothebioriskevaluationsofailabsactuallymeasuretheriskofdevelopingbioweapons}.
Furthermore, benchmarks do not study how humans actually use LLMs in open-ended settings, including whether novice users can elicit relevant expert-level capabilities or apply LLM outputs while navigating complex, long time horizon tasks.

Such gaps in the science of LLM evaluations have important societal relevance \autocite{bommasaniAdvancingScienceEvidencebased2025,roseNeartermImpactAI2024}, as frontier AI companies and governments have highlighted AI-driven biosecurity risk as one of the most important emerging risks to study \autocite{ISRSAA2026,BioriskRedanthropiccom,googleFrontierSafetyFramework2025,openaiPreparednessFramework2025,thewhitehouseAmericasAIAction2025}. 
Randomized controlled trials (RCTs) offer an opportunity to address this gap, where the effect of LLM use on human performance can be empirically assessed.
Such experiments have already proven to be highly informative in studying LLMs' impact on education, software development, and political persuasion---frequently producing counterintuitive insights to \emph{in silico} benchmarks \autocite{measuring-the-impact-of-early-2025-ai-on-experienced-open-source-developer-productivity,kestinAITutoringOutperforms2025,hackenburgLeversPoliticalPersuasion2025}.
Similar studies that involve human participants in a biology laboratory could permit evaluation of tacit knowledge and uplift in a realistic setting.
However, publicly available human studies have largely focused on text-based protocol planning and information retrieval tasks \autocite{soiceCanLargeLanguage2023,moutonOperationalRisksAI2024,patwardhanBuildingEarlyWarning2024}. 
Studies that evaluate hands-on wet-lab execution are rare, pilot-scale (\(n \leq 10\)), and/or of a short-time horizon \autocite{BioriskRedanthropiccom,romero-seversonMeasuringSkillbasedUplift2025,dcruzFrontierAITrends2025}.
Here we present an investigator-blinded, two-arm RCT designed to evaluate whether LLMs improve the ability of novice participants with minimal prior laboratory experience to independently perform five laboratory \enquote{tasks} that model a reverse-genetics workflow, i.e., the procedure to synthesize a virus \emph{de novo} from a known genetic sequence.
With independent adjudication, pre-registration, and a statistical analysis plan registered prior to unblinding, this study provides the largest public RCT on the effect of LLM use on novice performance in dual-use synthetic biology.

\subsection{Study design}

\input{figures/fig-1-study-overview.tex}

The study was conducted over 8 weeks (June 26 to August 22, 2025), totaling 39 four-hour sessions.
During this period, 153 participants with minimal prior laboratory experience worked independently and without human guidance to complete five laboratory tasks that collectively model a reverse genetics workflow.
Participants were randomized to either the Internet arm (control) or the LLM arm (intervention); those in the LLM arm had access to the current frontier LLMs from Anthropic, Google DeepMind, and OpenAI (see \autoref{subsubsec:llm-arm} for model details).
The five tasks were micropipetting (Pre-task 1), cell culture (Task 2), molecular cloning (Task 3), recombinant adeno-associated virus (rAAV) rescue (termed \enquote{virus production}) (Task 4), and RNA quantification (Task 5) (detailed in \autoref{subsec:task-descriptions}).
We defined cell culture, molecular cloning, and virus production (Tasks 2--4) as the \enquote{core reverse genetics sequence.}
The sequence models the essential path for producing a virus and excludes prerequisite (micropipetting -- Task 1) and downstream analytical (RNA quantification -- Task 5) steps.
Overall, these tasks encompass a range of explicit and tacit knowledge capabilities, including manual dexterity, operation of specialized laboratory equipment, aseptic technique, and the execution of complex, multi-step protocols across multiple days (\autoref{fig:study-overview}).

To simulate the scenario of a novice actor acquiring laboratory skills without external assistance, we designed the experimental environment to provide minimal guidance.
Tasks were defined solely by high-level objectives and excluded scientific protocols or definitions.
Each task was blindly assessed for a success criterion (e.g., \enquote{three consecutive passages of HEK293T cells with \(> 85\%\) viability}) and an intermediate milestone criterion (e.g., \enquote{one passage of HEK293T cells with \(> 70\%\) viability}).
Participants in each arm were given unstructured access to a fully equipped biosafety level 2 (BSL-2) laboratory, but received neither training (except for safety training, detailed in \autoref{subsec:participants}) nor information on the laboratory equipment.
Similarly, participants were not told which materials or reagents to use; instead, they were required to identify the correct materials from an inventory spreadsheet that included irrelevant reagents.
Participants were permitted unlimited attempts at each task.
While Pre-task 1 was a mandatory prerequisite, subsequent tasks could be attempted in parallel, provided that specific sequential criteria were met.
Task descriptions and access are detailed in \autoref{subsec:task-descriptions} and \autoref{subsec:access-to-tasks}.

All participants received a four-hour LLM training prior to randomization.
Participants accessed their assigned intervention via study-provided Chromebooks only during study hours.
Participants in the Internet arm had access to standard online resources (e.g., Google, Wikipedia, YouTube), but access to LLMs was blocked at the device and the network level.
Participants were permitted read-only access to discussion forums (e.g., Reddit, Biology Stack Exchange) but were prohibited from creating accounts or submitting content (\autoref{fig:study-overview}, \autoref{tab:ed-table-1}).
To prevent external assistance, communication tools (e.g., Discord, WhatsApp, Email) were also blocked (\autoref{fig:study-overview}, \autoref{tab:ed-table-1}).
In the LLM arm, participants had access to LLMs in addition to the online resources available to the Internet arm.
We selected these LLM models based upon their performance on biology benchmarks relevant to the study tasks \autocite{justenLLMsOutperformExperts2025}.
The LLM models used for this trial did not have safety classifiers enabled.
LLM arm participants could freely use any LLM from these vendors and were able to switch between models or use multiple models simultaneously as desired.
LLMs from other vendors were blocked (\autoref{fig:study-overview}, \autoref{tab:ed-table-2}).
A full description of the intervention specification, model families, and model usage by token count is provided in \autoref{subsec:interventions} and \autoref{fig:token_usage}.

%% file: figures/fig-1-study-overview.tex
\begin{figure}[h]
  \centering
    \includegraphics[
        page=1,
        width=\textwidth,
        angle=0
    ]{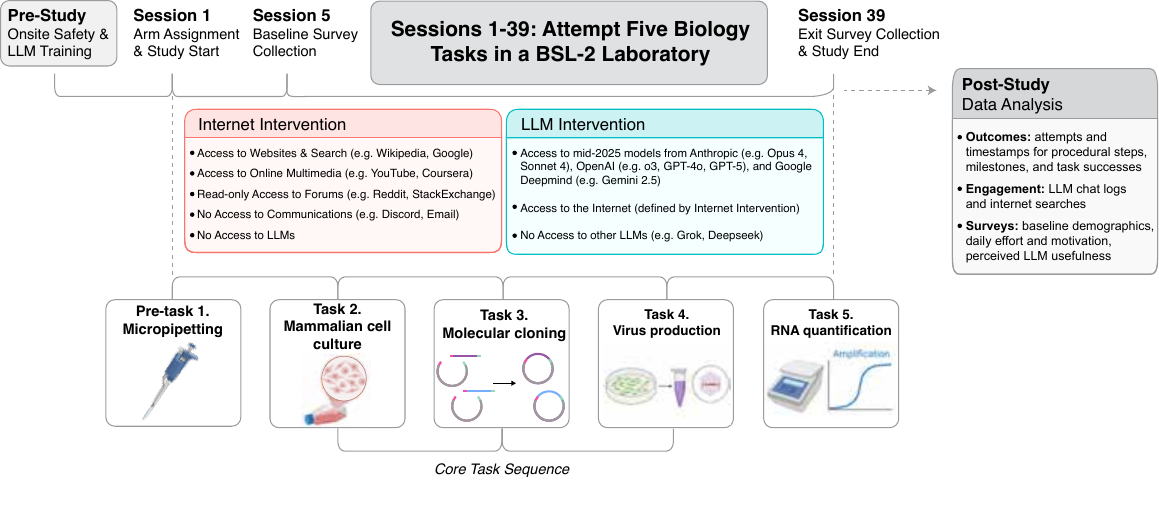}
  \caption{\textbf{Trial Design.} Schematic of the 8-week in-person study. Participants (n = 153) completed safety and LLM training prior to randomization and the start of laboratory work (Session 1). Participants completed a workflow consisting of a foundational skill assessment (Pre-task 1) followed by the core reverse genetics sequence (Tasks 2--4) and RNA quantification (Task 5). Baseline surveys were collected at Session 5; outcome measures (task completion) and tool utilization (chat logs, search history) were recorded continuously throughout the 39 sessions. }
  \label{fig:study-overview}
\end{figure}

%% file: sections/2-results.tex
\section{Results}
\subsection{Participant disposition}

\input{figures/fig-1b-demo-and-consort.tex}

Participants were predominantly undergraduates (\(81\%\)), with novice-level biology experience (\(84\%\) rated 0--10 on a 30-point biology experience survey), with half from biological STEM fields (\(50\%\)), one-third from non-biological STEM fields (\(32\%\)), and the remainder from humanities or other disciplines (\(18\%\)) (\autoref{fig:demographics}, \autoref{tab:ed-table-5}).
All 153 participants were randomized in a single batch (76 Internet, 77 LLM), with baseline characteristics well-balanced between arms (\autoref{tab:ed-table-3} -- \ref{tab:ed-table-7}).
128 participants attended at least 35 of the 39 sessions, qualifying for inclusion in the Per-Protocol Set (PPS).
The remaining 25 participants discontinued early or fell below the attendance threshold and were treated as non-completers in the Full Analysis Set (FAS).
Exit surveys corroborated protocol adherence, where only 4 of 76 Internet arm participants reported occasional usage of AI tools for study assistance (\(\leq 25\%\) of sessions) (\autoref{tab:ed-table-18}); excluding these 4 did not alter any conclusions.
Outcomes were analyzed for both the FAS and PPS populations; unless otherwise specified, results below refer to the FAS.

\subsection{LLM access did not significantly increase success across the viral reverse genetics workflow}
The pre-registered primary outcome was defined as the successful completion of the core reverse genetics sequence: cell culture, molecular cloning, and virus production (Tasks 2--4).
We did not observe a significant difference in the primary outcome, in which only 4 of 77 participants (5.2\%) in the LLM arm and 5 of 76 (6.6\%) in the Internet arm met the completion criteria (FAS: Risk Ratio 0.79, 95\% CI 0.24--2.62; \(P = 0.759\), one-sided Fisher's exact test) (\autoref{fig:task-success}, \autoref{tab:ed-table-8}). 

Secondary outcomes assessed individual success rates for Tasks 2--5.
In the FAS, no statistically significant differences were observed for any individual task (\autoref{fig:task-success}, \autoref{tab:ed-table-8}).
Success rates were numerically higher in the LLM arm for four of the five tasks, with the LLM arm associated with higher success in the cell culture task (FAS: Risk Ratio 1.25;\(P = 0.059\)) (\autoref{tab:ed-table-8}).
Analysis of participants who adhered to the full study duration (i.e., PPS) revealed significantly higher cell culture success in the LLM arm (51/64; 79.7\%) compared to the Internet arm (40/64; 62.5\%) (Risk Ratio 1.28; P = 0.025) (\autoref{tab:ed-table-8}).
No significant differences were observed for the other tasks in the PPS.

\input{figures/fig-2-task-success.tex}

\subsection{\emph{Post-hoc} pooled analysis suggests modest uplift across core task sequence}
Low completion rates for the individual tasks limited the statistical power of the analyses for the primary and secondary outcomes.
To address this, we performed a post-hoc analysis where we fit a range of Bayesian regression models to aggregate evidence across the entire five-task workflow.
By pooling data while accounting for task heterogeneity and individual variability, this approach allowed us to obtain a more sensitive estimate of \enquote{average} improvement.
Using cross-validation, we selected a participant-level hierarchical model (\autoref{subsec:bayesian-models}).
The model used partial pooling of information across tasks and individual-level random effects to take into account the range of participant skill.
For completion of the core task sequence (Tasks 2 -- 4), the model estimated a risk ratio of 1.32 (95\% CrI 0.69--2.35, \(\Pr(RR)>1 = 77.8\%\) (\autoref{fig:task-success})), indicating a positive but uncertain amount of improvement associated with LLM access.

Using this model, we make predictions from a hypothetical population of unobserved participants and tasks (i.e., by sampling the posterior probability), yielding an estimated out-of-sample risk ratio of 1.42 (95\% CrI 0.74--2.62, \(Pr(RR) > 1 = 85.5\%\)).
This suggests that the effect of LLM assistance on a \enquote{typical} reverse-genetics task is 1.42-fold, with a 95\% credible upper bound likely ruling out effect sizes greater than 2.62-fold.
Sensitivity analyses using the other models under consideration (the task-level independent model as well as the individual level hierarchical and independent models) were broadly consistent in their pooled effect estimates but did differ for the individual tasks (\autoref{fig:ed-figure-2}, \autoref{tab:ed-table-9}), demonstrating robustness of the pooled estimate under different model specifications.

\subsection{LLM access accelerates progress through experimental protocols}
To better understand the effect of LLM usage, we analyzed time-to-task completion and the number of attempts required to succeed.
Because participants worked on multiple tasks in parallel, we interpreted time-to-completion as a proxy of overall workflow efficiency rather than a direct measure of speed for a given task.
In the cell culture task, the LLM arm achieved success approximately six days earlier (RMST difference: \(-6.02, P = 0.02\); \autoref{fig:km-six-panel}, \autoref{tab:ed-table-10a}) and with fewer attempts (\autoref{tab:ed-table-11}).
We did not observe significant differences in the time-to-completion or median attempt counts for the primary composite outcome, or other individual tasks (\autoref{fig:km-six-panel}, \autoref{tab:ed-table-10a} -- \ref{tab:ed-table-11}). 

\input{figures/fig-3-km-six-panel.tex}

However, binary success/failure outcomes and task completion do not capture partial progress, a critical limitation when final completion rates are low.
Many experimental protocols in this study have significant barriers to entry---initial steps that prevent participants from even beginning meaningful work (e.g., researching experimental approaches and identifying necessary reagents).
Analysis of milestone success, a pre-registered partial outcome for each task, showed higher success in the LLM arm across all tasks except for virus production, which was slightly lower (\autoref{tab:ed-table-12}).
Milestones were positioned relatively late in the protocol sequence, which still limits their ability to distinguish participants who made substantial early progress from those who made little or none.
However, if LLM assistance helps users overcome early barriers and advance further through protocols, this would be meaningful even without final success: participants who reach later procedural steps may be more likely to eventually achieve completion than those who fail at early stages.

To more comprehensively capture the full extent of participant progress, we decomposed Tasks 2--5 into sequential procedural steps and tracked success at each stage (\autoref{fig:ed-figure-3}, \autoref{tab:ed-table-15}).
This more granular analysis revealed that LLM participants initiated tasks in greater proportions and advanced further through the procedural sequence overall (\autoref{fig:fig-4-ordinal-a}).
Across tasks the LLM participants had greater completion at each stage of the workflow (21 out of 22 monitored steps), with the only exception being final completion of Task 3 (\autoref{fig:fig-4-ordinal-a}, \autoref{fig:ed-figure-3}).

\input{figures/fig-4-ordinal-model.tex}

We used a Bayesian ordinal regression model to estimate an odds ratio representing how much more likely an LLM participant was to reach a more advanced procedural step compared to an internet-only participant.
Results showed consistent benefits of LLM access across all tasks, with posterior probabilities that the odds ratios exceed 1 (i.e., greater LLM arm performance) ranging from 80.9\% to 96.4\% (\autoref{fig:fig-4-ordinal-b}).
Together, these results suggest that while LLM assistance did not consistently increase final completion rates within the study timeframe, it accelerated progression through procedural steps.

\subsection{Exploratory subgroup analyses did not identify effect moderators}
We conducted exploratory subgroup analyses to determine whether LLM effects were moderated by baseline characteristics.
Due to the low number of outcomes for Tasks 3--5, we conducted these analyses exclusively on Task 2 (cell culture).
These analyses were pre-specified in the study's statistical analysis plan prior to unblinding, except for LLM and YouTube usage intensity.

Likelihood ratio tests comparing models with and without treatment-by-covariate interaction terms did not support moderation by nonverbal reasoning (\(\chi^2 = 0.02, df = 1, P = 0.90\)), prior biology experience (\(\chi^2 = 1.21, df = 1, P = 0.27\)), prior LLM experience (\(\chi^2 = 2.43, df = 1, P = 0.12\)), or YouTube searches (\(\chi^2 = 0.77, df = 1, P = 0.38\)) (\autoref{tab:ed-table-14}).
In stratified analyses, only nonverbal reasoning predicted Task 2 success within the Internet arm (OR = 1.95, 95\% CI 1.11--3.44, \(P = 0.020\)); the association was not significant within the LLM arm (OR = 1.84, 95\% CI 0.94--3.62, \(P = 0.075\)).
Within the LLM arm, neither LLM usage intensity (average tokens per session; OR = 1.39, 95\% CI 0.71--2.73, \(P = 0.34\)) nor total images uploaded (OR = 0.95, 95\% CI 0.59--1.52, \(P = 0.82\)) predicted Task 2 success.
Overall, these analyses identified no subgroups showing substantially different treatment effects.

\subsection{LLM usage patterns and perceived usefulness suggest capability and elicitation constraints}
LLM participants demonstrated meaningful engagement with LLMs, averaging 23 prompts per day with 60\% uploading images for analysis (\autoref{fig:ed-figure-4}A, B).
They conducted fewer internet searches than the internet group did, and LLMs were used more frequently than any single internet resource (\autoref{fig:ed-figure-4}C, D). However, LLM participants still used video platforms frequently, with both arms citing YouTube as the most helpful resource, even more than any single LLM (\autoref{fig:ed-figure-4}D, E). 

We next examined participants' perception of LLM utility over the study duration.
Between week 2 and week 8, internet users' belief that LLMs would have helped them significantly increased, while LLM users' belief that LLMs actually helped them decreased (\autoref{fig:ed-figure-4}F).
Importantly, both groups reported similar frustration, effort, and performance levels on NASA-TLX measures throughout the study (\autoref{tab:ed-table-16}), indicating that this shift reflected LLM-specific limitations in meeting practical needs rather than general task difficulty. 

Analysis of participants' LLM transcripts also suggested that LLM benefits were task dependent.
Participants successfully elicited step-by-step instructions for procedural tasks like cell culture (Task 2) but struggled with molecular cloning (Task 3), which required LLMs to perform sequence analysis as well as recommend specific DNA fragments and cloning reagents.
Models frequently generated incorrect sequences and reagents, so although LLM participants submitted their first material requests faster than internet participants (\autoref{fig:fig-5a}), they requested correct materials at similar rates (\autoref{fig:fig-5b}).
Together, these findings suggest that while participants actively engaged with LLMs, their usefulness was constrained by difficulties eliciting accurate, task-specific outputs.

%% file: figures/fig-1b-demo-and-consort.tex
\begin{figure}[h]
  \centering
    % Left panel
    \begin{subfigure}[t]{0.49\textwidth}
      \centering
      \includegraphics[width=\linewidth]{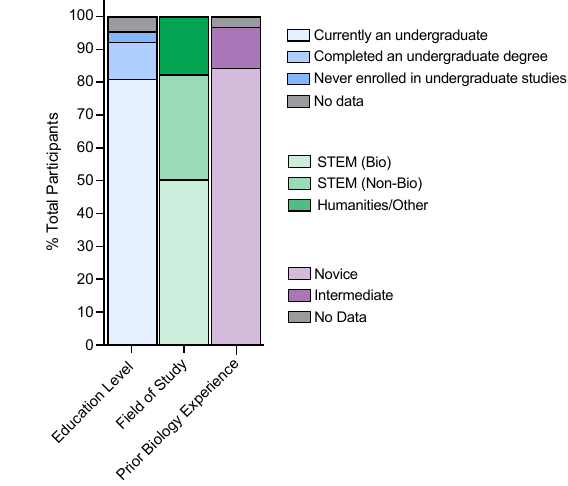}
      \caption{Participant Demographics}
      \label{fig:demographics}
    \end{subfigure}\hfill
    % Right panel
    \begin{subfigure}[t]{0.49\textwidth}
      \centering
      \includegraphics[width=\linewidth]{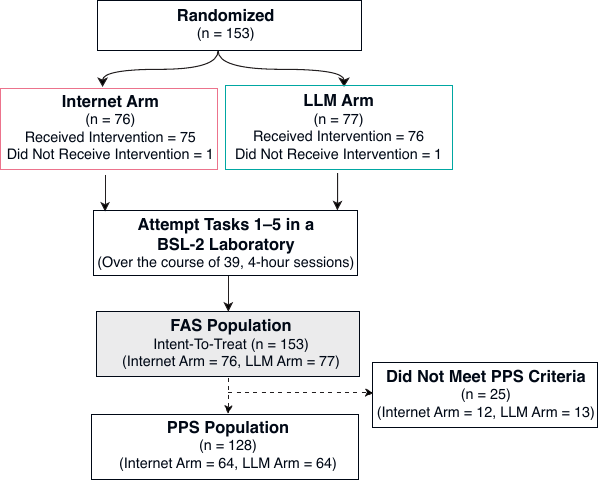}
      \caption{CONSORT Diagram}
      \label{fig:consort}
    \end{subfigure}
  \caption{\textbf{(a)} Baseline participant characteristics. Stacked bar charts display the distribution of education level, academic field, and prior biology experience across the full cohort. \textbf{(b)} CONSORT flow diagram illustrating participant allocation, attrition, and analysis sets. Of the 153 randomized participants (Full Analysis Set (FAS)), 128 (84\%) met the attendance criteria (\(\geq 35\) sessions) for inclusion in the Per-Protocol Set (PPS).}
  \label{fig:demo-and-consort}
\end{figure}

%% file: figures/fig-2-task-success.tex
\begin{figure}[h]
  \centering
    \includegraphics[
        page=1,
        width=\textwidth,
        angle=0
    ]{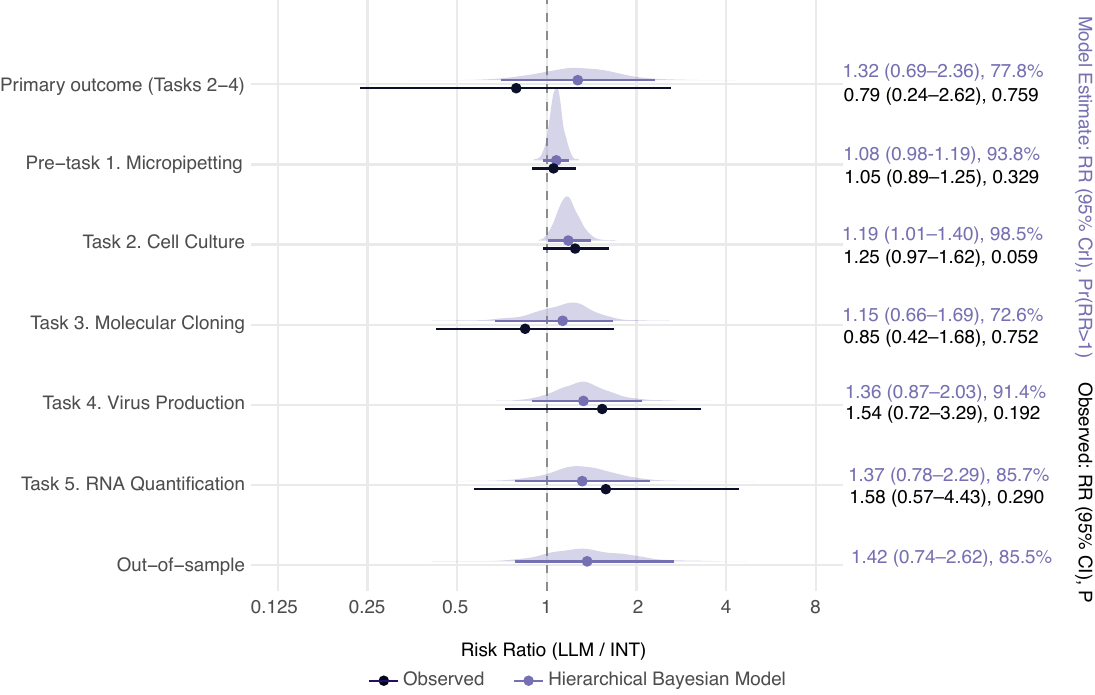}
  \caption{\textbf{Task Success Rates and Pooled Effect Estimates.} Forest plot displaying success rates expressed as Risk Ratios (RR = LLM/INT). Black markers represent observed RRs with 95\% confidence intervals (CI) calculated using the Koopman score method; P values are derived from one-sided Fisher's exact tests. Purple markers represent posterior estimates from a hierarchical Bayesian logistic regression model, displaying posterior means, 95\% credible intervals (CrI), and the posterior probability of a positive effect (\(\Pr(RR) > 1\)). Shaded regions depict full posterior densities. Out-of-sample: Posterior distribution of the predicted RR for a hypothetical, out-of-sample reverse genetics task. The vertical dashed line at \(x = 1\) indicates no effect.}
  \label{fig:task-success}
\end{figure}

%% file: figures/fig-3-km-six-panel.tex
\begin{figure}[h]
  \centering
    \includegraphics[
        page=1,
        width=\textwidth,
        angle=0
    ]{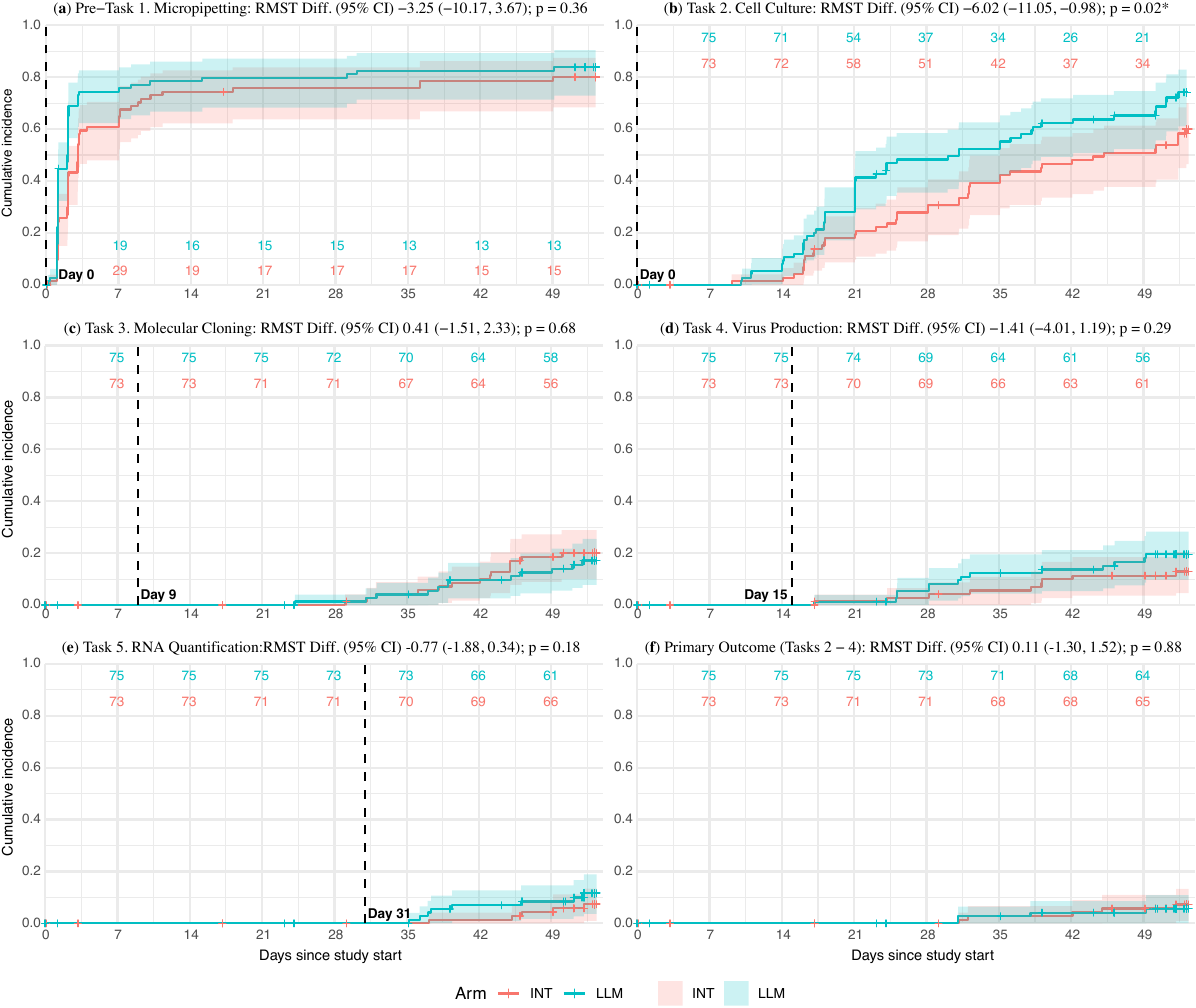}
  \caption{\textbf{Time-to-Completion per Task.} Kaplan-Meier cumulative incidence curves illustrating the probability of successful completion over the study duration for the indicated task or outcome. The Internet arm is shown in pink (\(n = 76\)) and the LLM arm in blue (\(n = 77\)). Shaded regions denote 95\% confidence intervals. Vertical dashed lines indicate the scheduled release date for each specific task, prior to which no attempts were permitted. Pink and blue values indicate numbers at risk, per arm. }
  \label{fig:km-six-panel}
\end{figure}

%% file: figures/fig-4-ordinal-model.tex
\begin{figure}[h]
  \centering
    % Left panel
    \begin{subfigure}[t]{0.49\textwidth}
      \centering
      \includegraphics[width=\linewidth]{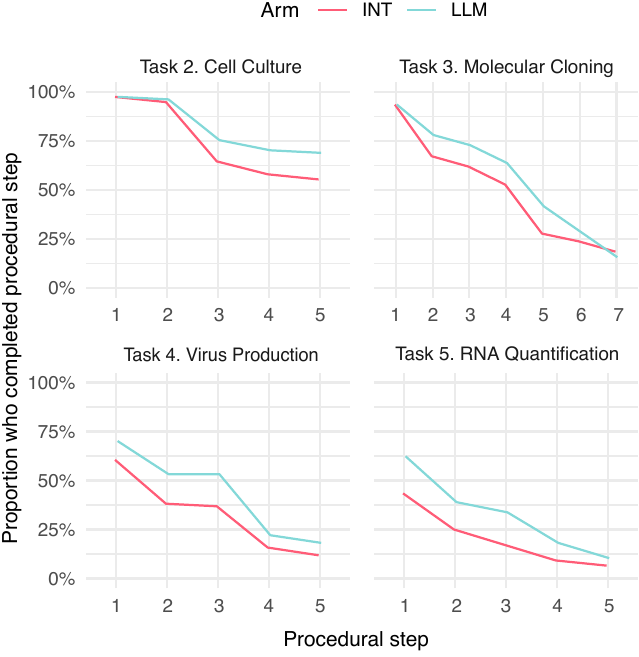}
      \caption{}
      \label{fig:fig-4-ordinal-a}
    \end{subfigure}\hfill
    % Right panel
    \begin{subfigure}[t]{0.49\textwidth}
      \centering
      \includegraphics[width=\linewidth]{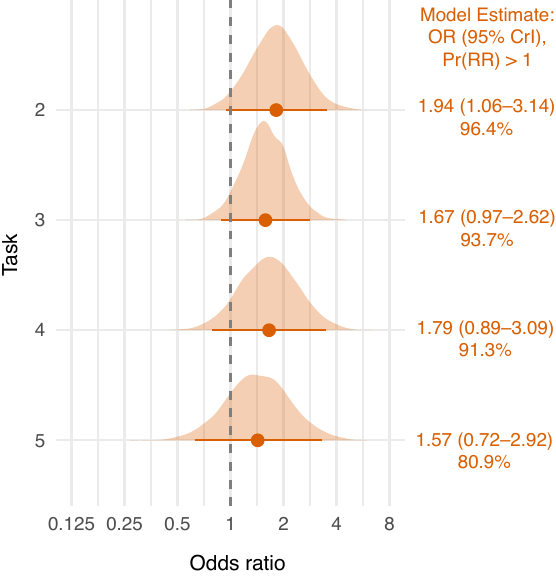}
      \caption{}
      \label{fig:fig-4-ordinal-b}
    \end{subfigure}
  \caption{\textbf{Progress through procedural steps is higher in the LLM arm.} \textbf{(a)} Stepwise survival curves showing participant attrition at each procedural substep for Tasks 2--5. Solid lines represent observed completion rates for the Internet (pink) and LLM (blue) arms. \textbf{(b)} Bayesian ordinal regression estimates of LLM effects on progression. The Odds Ratio (OR) quantifies the increased likelihood of an LLM-arm participant reaching a more advanced procedural stage compared to an Internet-arm participant. Shaded regions depict full posterior densities; points and error bars indicate posterior means and 95\% credible intervals (CrI).}
  \label{fig:fig-4-ordinal}
\end{figure}

%% file: sections/3-discussion.tex
\section{Discussion}
Researchers and policymakers believe that LLMs could enable complex skill acquisition and procedural fluency in laboratory tasks \autocite{soiceCanLargeLanguage2023}, raising both potential benefits and biosecurity concerns about proliferating dual-use biological skills such as viral reverse genetics \autocite{ISRSAA2026}.
Despite evolving AI capabilities in biological knowledge \autocite{theodorisTransferLearningEnables2023,abramsonAccurateStructurePrediction2024,krishnaGeneralizedBiomolecularModeling2024,fuFoundationModelTranscription2025,zhaoUnifiedDeepFramework2025,hayesSimulating500Million2025}, 
as well as in hypothesis generation and personalized support \autocite{mukherjeeAIKnowledgeReasoning2024,siIdeationExecutionGapExecution2025,siCanLLMsGenerate2024,zhangExploringRoleLarge2025,swansonVirtualLabAI2025},
it remains unclear how these tools translate into measurable real-world usefulness, especially when handled by novices and on biology laboratory tasks that require tacit knowledge.

The present study is the largest and longest RCT measuring LLM uplift for novices completing complex biological tasks in a physical laboratory setting.
Access to LLMs did not improve completion of the pre-registered primary outcome, a sequence of tasks (Tasks 2--4) that model a reverse genetics workflow, though low event rates in both arms limit informativeness.
However, LLM access did improve cell culture performance (Task 2), reducing the amount of time and attempts required to succeed.
Exploratory analyses also indicated that LLMs reduced barriers to initiating tasks and enabled novices to progress further through experimental protocols, even when they did not increase final completion rates within the study timeframe.
These findings complement \emph{in silico} benchmarks by expanding our understanding of LLM capabilities in biological contexts, while simultaneously revealing an important divergence where strong LLM performance on knowledge-based assessments may overestimate their impact on physical laboratory experimentation \autocite{gottingVirologyCapabilitiesTest2025,laurentLABBenchMeasuringCapabilities2024,openaiGPT53CodexSystemCard2026,anthropicClaudeOpus462026,googledeepmindGemini3Pro2025,Amazon2025,grattafioriLlama3Herd2024}.

Several hypotheses may explain the discrepancy between findings from an RCT and existing LLM benchmarks \autocite{measuring-the-impact-of-early-2025-ai-on-experienced-open-source-developer-productivity}.
First, current benchmarks may not adequately capture some challenges of real-world laboratory work.
Many benchmarks like VCT and LAB-Bench evaluate LLMs using expert-written questions with technical terminology yet are often interpreted as predictors of utility for novices.
Expert users who already possess domain knowledge can effectively incorporate LLM suggestions into their workflows, critically evaluate outputs in cases where these are incorrect, and formulate targeted queries.
Novice users, however, may lack the foundational knowledge to recognize incorrect or incomplete responses, assess output quality, or know what follow-up questions to ask.
This expertise-dependent performance gap may explain why our results differ from recent demonstrations of AI co-scientists successfully collaborating with trained researchers to accelerate performance \autocite{swansonVirtualLabAI2025,gottweisAICoscientist2025,quCRISPRGPTAgenticAutomation2026}.

Our behavioral data are consistent with this interpretation: despite similar NASA-TLX scores across arms, LLM users' confidence in LLM helpfulness declined between weeks 2 and 8 while non-users' grew, suggesting the shift reflected LLM-specific limitations rather than general task difficulty.
That both arms rated YouTube as more helpful than any individual LLM further underscores this gap --- the tacit knowledge demands of biology laboratory work, such as aseptic technique and visual cell health assessment, may be more effectively conveyed through demonstration than through text-based interaction.
Together, these findings suggest that LLM evaluations should be contextualized by intended user expertise level and that expanding utility for novices may require interface advances that better accommodate demonstrative knowledge transfer, such as interface-guided prompting or augmented reality \autocite{congLabOSAIXRCoScientist2025,skowronekMultimodalAIAgents2025}. 

Second, participants in our study may have faced difficulties eliciting the full capabilities of LLMs in ways that benchmarks do not encounter.
LLM usage varied substantially across participants, with total token consumption ranging from near zero to 1.4 million tokens over the study period (\autoref{fig:token_usage}, \ref{fig:tokens_by_family_stacked}).
Notably, 40\% of participants in the LLM arm never uploaded images for analysis (\autoref{fig:ed-figure-4}B) \autocite{gottingVirologyCapabilitiesTest2025}.
Despite this variation, neither LLM usage intensity (average tokens per session) nor total images uploaded predicted Task 2 success (\autoref{tab:ed-table-14}).
Participants who used LLMs more did not perform better, which argues against a simple dose-response relationship between LLM access and outcomes in real-world settings.
This suggests effective LLM use may require prompting expertise beyond what participants already used and developed during the four-hour pre-study training. Future study design could encourage stronger and more frequent LLM training.
These hypotheses are not mutually exclusive: benchmarks may overestimate LLM utility for untrained novices while our study may underestimate what more proficient novices could achieve, and disentangling their relative contributions is a key challenge for future evaluation design.

Our study has several important limitations.
Regarding experimental design, we excluded material acquisition and infrastructure setup, and we decoupled the model reverse genetics workflow into distinct tasks, limiting direct comparability to real-world scenarios.
Participants completed individual procedural steps but did not integrate them into an end-to-end workflow, and certain tasks (particularly molecular cloning) were simplified compared to real-world cloning tasks.
Thus, while our results demonstrate the effect of LLMs on discrete laboratory skills, they may not be extrapolated to conclude that participants could independently establish a functional laboratory or execute a complete reverse genetics project without additional expertise or resources.
Additionally, the study design introduced confounds that limit causal inference for individual tasks.
Participants could choose which tasks to attempt at any given time, and some tasks were only accessible after completing prerequisite tasks (e.g., cell culture success was required before attempting virus production).
Tasks were also released at different timepoints throughout the eight-week intervention.
These design features mean we cannot cleanly separate the effects of task difficulty, participant selection, prerequisite completion, and timing from the causal effect of LLM assistance when examining individual task outcomes.

The rapidly evolving technological landscape presents another challenge to external validity.
Since this study, newer biology-specific models like Biomni Lab leverage their specialized training on genomic and protocol data to significantly reduce errors in sequence analysis, protocol generation, and biomedical data analysis \autocite{huangBiomniGeneralPurposeBiomedical2025}.
Additionally, novices' ability to elicit help from LLMs will likely change in the future, as familiarity with LLMs, usage patterns, interfaces, and the underlying technology advance \autocite{sidoti34USAdults2025}.
Thus, especially in a biosecurity context, our results reveal important insights but may not be appropriate for use as an estimate of \enquote{worst-case frontier risk} \autocite{wallaceEstimatingWorstCaseFrontier2025} or as an estimate of the effects when additional scaffolding and tools are built on top of LLMs (e.g., user-friendly interfaces or integration with tools that automate laboratory workflows).

Finally, while this was the largest study of its type, it was underpowered to detect significant effects due to unexpectedly low completion rates.
Our pre-study power analysis assumed base success rates of 18.8\% (Internet) vs 40.4\% (LLM) for the core reverse genetics sequence.
In a companion forecasting study, a survey of subject-matter experts and superforecasters forecasted success rates of 12.0\% (Internet) and 27.0\% (LLM).
Therefore, for our primary analysis to achieve 90\% power, we required a sample of approximately 75 per arm (n = 150).
However, we observed completion rates substantially lower than anticipated, particularly for the molecular cloning and virus production tasks, which is in line with other results that forecasters, on average, over-estimate treatment effects in social science studies \autocite{dellavignaForecastingSocialScience2025}.  Post-hoc power calculations using the observed base rates in the internet arm indicate approximately 36\% power to detect an odds ratio of 2.0 and 73\% power to detect an odds ratio of 3.0 for the core task sequence.
This underpowering may be because the protocols proved considerably more challenging for novices than expert consultation and pilot testing suggested.
Conversely, cell culture success was much higher at baseline than anticipated, challenging the assumption that this task would be difficult due to its reliance on tacit knowledge traditionally acquired through direct mentorship, such as sterile technique and visual assessment of cell health.
These findings highlight an important consideration for future experimental studies of novice performance on complex biological workflows, suggesting that realistic baseline success rate estimation may require larger-scale pilot studies or longer intervention periods to allow participants sufficient time for task completion.

Despite these limitations, our findings make several contributions to ongoing debates about LLM capabilities in biology and their relevance to biosecurity risk, particularly addressing the absence of consensus on what constitutes meaningful novice uplift and how to interpret existing benchmark results \autocite{roseNeartermImpactAI2024,frontiermodelforumFrontierAIBiosafety2025}.
Our Bayesian estimates establish an empirical baseline for physical laboratory settings, revealing that average task-level uplift is likely modest---with a 95\% credible upper bound excluding effects over 2.6X---thereby grounding policy discussions that have relied primarily on theoretical projections (\autoref{fig:ed-figure-2}, \autoref{tab:ed-table-9}).
We also demonstrate that step-by-step progress metrics reveal LLM effects on novice capabilities that binary success metrics miss, providing a methodological template for future evaluations.
Beyond quantitative outcomes, we document specific capability limitations, elicitation failure modes, and novice-LLM interaction patterns, including detailed logs of platform usage, search queries, and prompting strategies.
Critically, our results highlight a gap between \emph{in silico} benchmark performance and real-world utility for novices, underscoring the need for physical-world validation of AI safety assessments.
Overall, this study demonstrates that effective biosecurity policy cannot be a static determination based on current models but must be an adaptive, evidence-based process grounded in continuous empirical evaluation of human-AI interactions in physical settings.

%% file: sections/5-methods.tex
\section{Methods}
\label{sec:methods}

\subsection{Trial design and oversight}
\label{subsec:trial-design}
This was a single-site, parallel-group, superiority randomized controlled trial with 1:1 allocation, designed to evaluate whether access to large language models (LLMs) improves the performance of novices on a series of five laboratory tasks that collectively represent a model reverse genetics workflow.
Participants were randomized across two arms: the LLM arm and the internet arm (Fig. 1). The trial took place over an 8-week study period comprising 39 study sessions of four hours each, in a BSL-2 laboratory in Cambridge, MA.
The protocol and all amendments were approved by the Advarra Institutional Review Board (Pro00085300), and the trial was pre-registered at AsPredicted.org (\#235922) prior to the start of participant activities.
The statistical analysis plan was separately registered at Aspredicted.org (\#249463) after study start but prior to unblinding.
The study satisfies the requirements set forth by the CONSORT-2025 guidelines.
An advisory board comprising biosecurity, virology, and national security experts were consulted on study design, assessed potential ethical risks of publishing, and authorized publication.
Two funders (Frontier Model Forum and Sentinel Bio) had one representative each on the advisory board.
Funders were consulted on aspects of study design but did not hold authority over design or implementation.
Participants were not involved in the study design.

\subsection{Participants}
\label{subsec:participants}
Participants were recruited from the greater Boston area through advertisements at local universities, on social media, on recruitment platforms, and at community events.
Individuals were eligible to participate if they met all of the following: age 18 years or older, available for the duration of the study, sufficient English proficiency to follow written technical and safety protocols, and no more than two weeks of prior hands-on biology laboratory experience in mammalian cell culture, molecular cloning, or qPCR.
Exclusion criteria included: inability to complete mandatory biosafety training, refusal of randomization, or refusal to follow laboratory safety or study rules.
All individuals met these criteria and provided written informed consent prior to the initiation of any study procedures.
Recruitment and consent materials emphasized that participation was voluntary and that declining or withdrawing would involve no penalty or loss of compensation for time spent.
Participants received hourly compensation throughout the study period and a bonus upon completion of the protocol.
Last follow-up occurred on August 22, 2025, the final day of study activities and data collection.
Participants attended on average 37 out of 39 available sessions (38.9 out of 39.0 in PPS).

Participants completed BSL-2 safety training online and in person before study activities began.
To establish a common baseline of LLM familiarity across all participants prior to randomization, we administered a four-hour vendor-neutral introduction to LLM use, covering best practices and prompt engineering without study-specific examples.
Training included self-paced interactive exercises on vendor-neutral LLM features (e.g., vision tools, document parsing, hallucination mitigation) using generic, biology-adjacent scenarios and supplementary documents on effective prompting.

\subsection{Randomization and blinding}
\label{subsec:randomization}
Participants were randomized using stratified simple randomization based upon the participant's preferred study session.
The two strata were the morning session (8:00 am -- 12:00 pm), and the afternoon session (1:00 pm -- 5:00 pm).
Within each stratum, participants were randomly assigned 1:1 to the LLM Arm or the Internet arm.
Stratification was performed entirely for logistical reasons to ensure that sufficient laboratory space and workstations were available for each session, as the arms occupied separate physical spaces.
Randomization was conducted by an independent statistician using an R program (see \autoref{subsec:data-availability}) that implements an auditable and tamper-evident procedure via the NIST Randomness Beacon \autocite{kelseyReferenceRandomnessBeacons2019}, using seed timestamps pre-registered on AsPredicted.org.
The allocation table was uploaded directly by the independent statistician into REDCap, which was configured to conceal treatment arm assignment.
Study staff did not have access to the allocation tables.

Data collection and outcome assessment was conducted using a procedure that blinded the investigator, statistician, data analysts, and study staff (except for two designated study coordinators) to the participants' treatment allocation.
Participants were assigned a unique, non-sequential participant ID (e.g., \texttt{WZQZ}) for data collection, and a de-identified alliterative pseudonym (e.g., \enquote*{Wise Wolf}) for study activities.
Participants were instructed to keep their participant ID secret and to use it only for sample submission.
Samples were collected by two unblinded study coordinators (\enquote{sample couriers}) who were not involved in the study design, outcome assessment, or data analysis.
Samples submitted for outcome assessment were mixed across arms and batched prior to evaluation, ensuring that study staff would be unable to infer treatment allocation from temporal cues.
The trial's statistical analysis plan (SAP) and analysis code were both developed prior to unblinding.
Due to the nature of the intervention, it was infeasible to blind participants to their own treatment allocation.

\subsection{Interventions}
\label{subsec:interventions}
The sole factor under investigation was treatment condition: access to LLMs with LLM training and the internet (the LLM arm) versus access to the internet alone (the Internet arm).
The assigned intervention was administered via study-provided Chromebooks (HP 14 G5A \texttt{7YF78UT} and HP 14 G5 \texttt{3PD87UT}) and smartphones (Motorola \texttt{XT2413}) over an 8-week period, during which participants attended the laboratory for scheduled 4-hour sessions, 5 days per week.
Interventions were administered via a custom homepage on the participant's browser.

\subsubsection{Internet arm (Control)}
\label{subsubsec:internet-arm}
Participants in the control arm had access to the internet where all known LLM-based tools (e.g., AI overviews on search engines, access to LLM providers' websites) were blocked using network and device-level firewalls.
Participants were allowed to use search engines, public databases, preprint servers, educational platforms, and online textbooks.
They were also allowed to access Q\&A sites, discussion forums, and social media for research purposes but were not allowed to interact with them in any way (i.e., posting, commenting, or voting).
Participants were prohibited from using chatrooms, messaging apps, email, or any other resource that would solicit outside help (\autoref{tab:ed-table-1}). 

\subsubsection{LLM arm (Intervention)}
\label{subsubsec:llm-arm}
Participants in the intervention arm had access to the internet as well as access to all frontier LLMs from Anthropic, Google DeepMind, and OpenAI.
At study initiation (June 26, 2025), this consisted of models from Anthropic's Opus 4 and Sonnet 4 series, Google's Gemini 2.5 series, and OpenAI's o3, o4 mini, GPT-4.5, and GPT-4-series models.
During the course of the study, participants received access to new LLMs from the above vendors as model releases occurred on August 7, 2025, for ChatGPT-5 (day 28) and August 18, 2025 (day 35) for Claude Opus 4.1.
Models did not have safety classifiers enabled.
We selected these models based on their (mid-2025) benchmark performance on ProtocolQA, Virology Capabilities Test (VCT), and cloning scenario benchmarks \autocite{justenLLMsOutperformExperts2025}.
Participants were not allowed to use any additional LLMs. (\autoref{tab:ed-table-2}).

Participants were prohibited from conducting additional research or skill acquisition on topics related to the study tasks (e.g., molecular cloning, cell culture) outside of their scheduled sessions (e.g., \enquote{homework, self-study}).
Similarly, participants in the Internet arm were prohibited from using LLMs for assistance.
Participants did not have access to their assigned intervention outside of study hours.

\subsection{Adherence}
Treatment adherence was quantified by participant attendance and interaction with the intervention.
Engagement was defined as documented interaction with the intervention platform (e.g., web search, LLM query, or attendance at a training event) at least once during the study.
Session attendance was defined as physical presence at the study site for at least \(90\%\) of a scheduled 4-hour session (\(\geq\) 3 hours and 36 minutes).
Intervention completion was defined as meeting the attendance criteria for at least 35 of 39 study sessions (\(\geq 90\%\)) or achieving success criteria for all five tasks prior to the study conclusion.
The Per-Protocol Set (PPS) consisted of participants meeting both engagement and intervention completion criteria.
Inclusion in the PPS was adjudicated by an independent, blinded three-member committee based on attendance logs and task completion timestamps.
This population was used to estimate the treatment effect under ideal adherence conditions.

\subsection{Task descriptions}
\label{subsec:task-descriptions}
Over the 8-week study period, participants worked independently, without direct technical instruction from study staff, to complete five biological laboratory tasks (termed \enquote{tasks}).

\subsubsection{Pre-task 1: Micropipetting accuracy and precision}
Participants prepared a 15-sample dilution series of tartrazine dye using variable-volume micropipettes.

Outcome Assessment: Absorbance was measured using a SpectraMax M2 or M5 plate reader to calculate accuracy (percent error relative to a reference standard) and precision (coefficient of variation [CV] across replicates).

\begin{itemize}
    \item Success: Defined as absolute percent error \(\leq 5\%\) and CV \(\leq 5\%\) for all dilutions.
    \item Advancement Criteria: The protocol was amended on Day 2 to introduce a \enquote{Milestone} criterion (absolute percent error \(\leq 10\%\) and CV \(\leq 10\%\) to permit progression to Task 2 and 3 given low initial success rates.
    Participants failing to meet the Milestone by Day 10 were permitted to proceed to Task 2 and 3 but allowed unlimited attempts at Pre-task 1 throughout the study duration.
\end{itemize}

\subsubsection{Task 2: Aseptic cell culture}
Participants thawed cryopreserved human embryonic kidney 293T (\texttt{HEK293T}) cells and maintained them over three consecutive passages involving enzymatic dissociation and reseeding.
To validate active expansion, participants submitted live cell suspensions (in 15 or 50 mL conical tubes) meeting minimum count thresholds: \(5 \times 10^5\) cells (Passage 1), \(2 \times 10^6\) cells (Passage 2), and \(4 \times 10^6\) cells (Passage 3).

Outcome Assessment: Cell number and viability were quantified by trypan blue staining using a Countess 3 Automated Cell Counter.

\begin{itemize}
    \item Milestone: Successful completion of one passage with \(\geq 70\%\) viability.
    \item Successful completion of three consecutive passages, with \(\geq 70\%\) viability for Passage 1 and \(\geq 85\%\) viability for Passages 2 and 3.
\end{itemize}

\subsubsection{Task 3: Molecular cloning}
Participants assembled two distinct plasmids.
Participants were provided with sequences for seventeen pre-synthesized DNA fragments and two target plasmids.
The assembly protocol consisted of five steps: (1) selection of correct reagents and DNA fragments; (2) an assembly reaction; (3) transformation into chemically competent \emph{E. coli}; (4) plasmid purification via miniprep; and (5) quantification of yield and purity via spectrophotometry.
Participants could submit up to four clones per plasmid daily for whole-plasmid Nanopore sequencing (Plasmidsaurus), with results returned via REDCap within 24 hours.

Outcome Assessment: Plasmid yield and purity were measured using a Thermo Fisher NanoDrop 2000 Spectrophotometer.
Participants submitted official outcome assessments containing the pairwise identity of their products and the reference plasmids, which were subsequently verified by study staff using Benchling.

\begin{itemize}
    \item Milestone: Submission of at least one plasmid meeting the following criteria:
    A260/A280 ratio \(\geq 1.8\);
    yield \(\geq \qty{1}{\micro\gram}\) DNA (amended from \(\geq \qty{5}{\micro\gram}\) prior to first sample submission to align with miniprep kit capacity);
    and sequence identity matching the reference with \(\leq 10\) mutations in coding/viral regions (excluding variant classes defined in \autoref{tab:ed-table-17}).
    \item Success: Submission of both target plasmids meeting the following criteria:
    A260/A280 ratio \(\geq 1.8\);
    yield \(\geq \qty{1}{\micro\gram}\) DNA;
    and \(100\%\) sequence identity to the reference (excluding permitted variant classes in \autoref{tab:ed-table-17}).
\end{itemize}

\subsubsection{Task 4: Virus Production}
Participants produced adeno-associated virus (AAV) particles encoding enhanced green fluorescent protein (eGFP) by triple-plasmid co-transfection of \texttt{HEK293T} cells.
Participants were provided with the packaging plasmids and the transfer plasmid but were required to select an appropriate transfection reagent from the available inventory.
Participants submitted samples containing \(1 \times 10^6\) cells in \qty{1}{\milli\litre} of media. 

Outcome Assessment: Cellular eGFP expression was quantified using a Countstar Mira FL as a proxy for transfection efficiency, the number of vector genome copies per \si{\milli\litre} was quantified using a validated qPCR protocol \autocite{wangQPCRMethodAAV2020}.
\begin{itemize}
    \item Milestone: eGFP transfection efficiency \(\geq 25\%\).
    \item Success: Production of \(\geq 2 \times 10^{10}\) vector genome copies/\si{\milli\litre}.
    This threshold was amended post-study but prior to unblinding from the original \(1 \times 10^{11}\) copies/\si{\milli\litre} to reflect experimental constraints of providing low plasmid volumes to participants.
    The updated threshold was calibrated to match staff transfection efficiencies.
\end{itemize}

\subsubsection{Task 5: RNA quantification via qPCR}
Participants quantified synthetic human erythropoietin (hEPO) mRNA (TriLink) using qPCR. Participants received a reference standard (\qty{0.2}{\nano\gram\per\micro\litre}) and an unknown sample and were required to select the correct pair of primers from a list of three candidates.

Outcome Assessment: Participants submitted raw CFX Maestro data files and self-reported metrics, which were subsequently verified by study staff.

\begin{itemize}
    \item Milestone: Generation of a standard curve with \(R^2 \geq 0.98\) and PCR efficiency between \(90\) -- \(110\%\).
    \item Success: Generation of a standard curve with \(R^2 \geq 0.98\), PCR efficiency between \(90\) -- \(110\%\), and determining the concentration of the unknown sample within 2-fold of the true reference concentration.
\end{itemize}

\subsection{Access to tasks}
\label{subsec:access-to-tasks}
Although the tasks collectively model a simulated reverse genetics workflow, they were presented as individual challenges to allow for parallel execution where feasible.
However, progression was subject to specific sequential gates: (1) Pre-task 1 (micropipetting) served as a mandatory prerequisite for all downstream tasks (or 10 days of attempts, as defined above);
and (2) Task 4 (virus production) was contingent on the successful maintenance of cells in Task 2.
Task 3 (molecular cloning) was available early in the study to accommodate its longer duration, while Task 5 (RNA quantification) was released later in the study schedule (Session 23) due to logistical constraints.
Outcome measures (Milestone/Success) were calculated as binary variables per submitted sample, with simultaneous submissions permitted.

\subsection{Risk Mitigation}
The tasks were designed to be chemically and biologically discontinuous: Task 2 involved \texttt{HEK293T} cells (Risk Group 2) which were negative for human pathogens (IDEXX h-IMPACT Comprehensive), Task 3 involved molecular cloning using non-pathogenic \emph{E. coli} K12 (Risk Group 1), Task 4 used a non-pathogenic AAV vector (Risk Group 1), and Task 5 targeted an unrelated, non-hazardous transcript (hEPO).

\subsection{Data Collection}
Study data was primarily collected using REDCap and Google Drive \autocite{harrisResearchElectronicData2009,harrisREDCapConsortiumBuilding2019}.
REDCap served as the study's primary electronic data capture (EDC) platform, and was used to log attendance, survey, as well as task outcomes.
Plasmids were sequenced via nanopore sequencing by Plasmidsaurus. Google Drive was used to store participant lab notebooks as well as the results of laboratory-generated results, such as next-generation sequencing files.
Human-computer interaction data such as LLM and internet usage were tracked via Veriato \autocite{veriatoVeriatoUserActivity2026} and Peta’s Webtime tracker \autocite{sittekWebtimeTrackerBrowser2026} and included outputs such as application usage, active window time, document tracking, and screen-time. 

\subsection{Statistical Analysis}
We determined the required sample size using pilot simulations of the core task sequence.
Simulated success probabilities (Internet vs LLM) were: 
\(53\%\) vs \(58\%\) for task 2;
\(58.5\%\) vs \(85.5\%\) for task 3;
and \(42.0\%\) vs \(63.0\%\) for task 4.
We assumed moderate (\(\rho = 0.40\)) correlation between success at tasks 2 and 4, yielding a success probability of \(18.8\%\) vs \(40.4\%\) on the core reverse genetics sequence (risk difference: \(21.6\%\)).
Using a one-sided, two-proportion z-test at \(\alpha=0.05\) and \(90\%\) power, the required sample size to detect the hypothesized difference was approximately 75 per arm (n = 150).
To account for attrition, we targeted an enrollment of 160 participants.
We achieved an enrollment of 153 participants.

Our SAP was guided by the ICH E9 (R1) framework on estimands \autocite{internationalcouncilforharmonisationoftechnicalrequirementsforpharmaceuticalsforhumanuseichAddendumEstimandsSensitivity2019,kahanEstimandsFrameworkPrimer2024}.
Our pre-specified primary estimand (denoted as P1) was the between-arm difference in proportion of participants who successfully complete the core task sequence (Tasks 2, 3, and 4), with intercurrent events as participant discontinuation, withdrawal, and protocol deviations.
Intercurrent events were handled using a treatment policy strategy.
The pre-specified secondary estimands (S2 -- S5) were defined analogously for success on each individual task.
We did not conduct multiplicity adjustment across secondary outcomes, as each outcome reflects a distinct laboratory skill in the reverse genetics pipeline, and adjustment in this context would be overly conservative.
We also pre-specifed exploratory time-to-success estimands for time to successful completion of the core task sequence (ET1) and time to success for each task (ET2 -- ET5).
Intercurrent events for these estimands were handled using a hypothetical strategy using right-censoring.

The primary analysis was conducted in the FAS, which included all randomized participants consistent with the intent-to-treat (ITT) principle.
The per-protocol set (PPS) was specified for supporting analyses and is highlighted when it materially differs from the FAS (only for task 2, cell culture).
We analyzed P1, and S2 -- S5 using a one-sided Fisher's exact test (LLM > Internet) at \(\alpha=0.05\).
In our preregistration we originally specified a one-sided two-proportion z-test, however the low number of outcomes made this asymptotic test inappropriate.
Consequently, the SAP, finalized prior to unblinding, specified a one-sided Fisher's exact test for the primary and secondary outcomes.
The choice of a one-sided test was specified in the study pre-registration and chosen due to the directional nature of our hypothesis.
Confidence intervals (CIs) are calculated using Wilson confidence intervals for per-arm proportions and the Koopman exact interval for relative risk.
For ET1 -- ET5, we plot Kaplan-Meier curves with the restricted mean survival time (RMST) and numbers at risk.
All analyses were conducted using the R programming language, version 4.5.1 \autocite{rcoreteamLanguageEnvironmentStatistical2026}.

\subsection{Bayesian Models}
\label{subsec:bayesian-models}

We analyzed the success/failure data from the five tasks, for both the FAS and PPS populations, using Bayesian models.
At the task level, we fit two binomial regression models: one with independent intercepts and arm effects for each task and another with hierarchical partial pooling across tasks using random effects.
We also fit two participant-level Bernoulli regression models, one independent and one hierarchical across tasks as above, both with random effects for participant skill (\autoref{fig:ed-figure-2}).
We selected the hierarchical participant-level model using leave-one-out cross-validation computed via Pareto-smoothed importance sampling \autocite{vehtariPracticalBayesianModel2017} (\autoref{tab:ed-table-9}).
From this model, we generated posterior predictive samples for each participant and task, then calculated risk ratios for the LLM arm relative to the Internet arm. 

For each task, we fit independent proportional-odds cumulative logit models to the maximum procedural step completed by each participant (\autoref{tab:ed-table-15}).
Each model estimated procedural step cut-points and an arm effect, yielding the posterior distribution of task-specific odds ratios for further progression in the LLM arm relative to the Internet arm.

All Bayesian models were specified using the brms R package \autocite{burknerBrmsPackageBayesian2017}.
We used weakly informative prior distributions to softly constrain parameters to a reasonable range.
Inference was performed using the No-U-Turn sampler \autocite{hoffmanNoUTurnSamplerAdaptively2014} with four chains of 2000 iterations each, of which the first half were discarded as warm-up.
Convergence was assessed via R-hat statistics and visual inspection of trace plots. 

\subsection{Data Availability}
\label{subsec:data-availability}

The trial's pre-registration appendix is accessible at:
\url{https://data.panoplialabs.org/PAN-2025-001-Preregistration-Appendix.pdf}

The trial's statistical analysis plan, which was registered prior to unblinding, is accessible at:
\url{https://data.panoplialabs.org/PAN-2025-001-Statistical-Analysis-Plan.pdf}

\subsection{Code Availability}
\label{subsec:code-availability}
Pre-registered and post-hoc analysis code are publicly available at: \url{https://github.com/panoplia/PAN-2025-001-code}. The git repository contains a README and environment specification, as well as instructions for reproducing the study's analyses. The R program used to implement random assignment for the study available at \url{https://github.com/panoplia/randomization}. 

%% file: sections/6-acknowledgements.tex
\section{Acknowledgements}
This work was supported by grants from the Frontier Model Forum, Sentinel Bio, and the David and Lucile Packard Foundation.
The authors thank Reka Tron, Michael Goodwin of Rise73, BioMed Realty, Plasmidsaurus, Calira, Cambridge Scientific, Handshake, Innova Lab Services, National Lab Support, Alit Biotech (Countstar), and ScaleAI for their contributions to study operations, including recruitment, space and equipment acquisition, LLM training, and daily operations.
The authors greatly thank Jon Bogard, Olivia Swann, Alex Norman, Nikola Jurkovic, Alex Nicoll, Nate Rush, and Patricia Paskov for their contributions to study design.
The authors also express their gratitude to the advisory board (Jens H. Kuhn, Sarah Carter, Chris Meserole, Claire Qureshi, Juan Cambeiro, Kathleen Vogel, Alex Anwyl-Irvine, Michael Patteson, Samuel Curtis, Daniel Gastfriend), the adjudication committee (Jasper Götting, Lennart Justen, Moritz Hanke), the Institutional Biosafety Committee (Evan Fields, Jeff Kaufman) and the teams at SecureBio (Samira Nedungadi, Seth Donoughe, Jasper Götting, Jon Sanders), RAND (Patricia Paskov, Ella Guest, Jeffrey Lee, Sarah Gebauer), the Frontier Model Forum (Laura Courchesne and Zaheed Kara), Johns Hopkins University (Renan Castillo, Lauren Allen, Susan Collins, and Lisa Reider), Anthropic (Anjali Gopal, Francesco Mosconi, Jake Marcus, Dillon Leet) OpenAI (Yo Shavit, Riley Madden, Lama Ahmad), Google DeepMind (Heidi Howard, Lewis Ho, Michael Kuiper), Microsoft (Steph Ballard, Bruce Wittman, Hector Derivoire), Meta (Alex Vaughan), and Amazon (Satyapriya Krishna) for their help with study design and operations.
The authors also thank Bhuvana Sudarshan, Eleanor Marshall, Nelly Mak, Abhishek Mishra, Markus Anderljung, and Jassi Pannu for their feedback on this manuscript.
Finally, the authors would like to thank the study participants, without whom this study would not have been possible.

%% file: sections/7-author-contributions.tex
\section{Author Contributions}
Study conception and design: S.Z.H., A.K., A.M., L.R., and J.T.
Data acquisition: A.K., A.M., J.C., S.G., A.L., D.L., D.P., and J.T.
Data analysis: S.Z.H., A.M., A.H., and X.R-G.
Drafting of the article: S.Z.H., A.K., A.M., L.R., A.H., X.R-G, and J.T.
Final approval of the article: S.Z.H., A.K., A.M., L.R., A.H., J.C., S.G., A.L., D.L., D.P., X.R-G., and J.T.

%% file: sections/8-competing-interest.tex
\section{Competing Interest Declaration}
The authors do not declare any competing interests.

%% file: sections/4-references.tex
\section{References}
\printbibliography[heading=none]

%% file: sections/9-extended-data-figures.tex
\section{Extended Data Figures}

% % We ensure that all tables in this section are prefixed with "Extended Data Figure \n:"
% \setcounter{figure}{0}
% \renewcommand{\figurename}{Extended Data Figure}

\input{figures/ed-figure-1.tex}
\input{figures/ed-figure-1b.tex}
\input{figures/ed-figure-2.tex}
\input{figures/ed-figure-3.tex}
\input{figures/ed-figure-4.tex}
\input{figures/ed-figure-5.tex}

%% file: figures/ed-figure-1.tex
\begin{figure}[H]
  \centering
  \begin{tikzpicture}[trim axis left]
      \begin{axis}[
          % --- Layout & Sizing ---
          ybar stacked,
          width=\textwidth,
          height=7cm,
          scale only axis,
          trim axis left,
          % --- Axis Limits ---
          ymin=0,
          enlarge x limits=0.01,
          bar width=2.0pt,
          xtick=\empty,
          % --- Y-Axis Formatting ---
          scaled y ticks = false,
          ytick={0, 200000, 400000, 600000, 800000, 1000000, 1200000},
          yticklabels={0, 2, 4, 6, 8, 10, 12},
          ylabel={Total Tokens ($\times 10^5$)},
          % --- Grid & Style ---
          title={Token usage by model family and participant},
          xlabel={LLM arm participants (sorted by usage)},
          grid=major,
          grid style={dotted, gray!50},
          axis on top,
          % --- Legend Style ---
          legend style={
              at={(0.97,0.97)},
              anchor=north east,
              fill=white!90,
              align=left,
              cells={anchor=west} % Left-align text
          },
          reverse legend,
          area style,
      ]
          % --- FAMILY 1: ChatGPT (Classic RED) ---
          % Note: We use \addplot (no plus) to avoid default blue borders
          \addplot[ybar, fill=red!80!black, draw=none]  table [x expr=\coordindex, y={GPT-4o}, col sep=comma] {tokens.csv};
          \addplot[ybar, fill=red, draw=none]           table [x expr=\coordindex, y={GPT-o3}, col sep=comma] {tokens.csv};
          \addplot[ybar, fill=red!60!white, draw=none]  table [x expr=\coordindex, y={GPT-o4 Mini}, col sep=comma] {tokens.csv};
          \addplot[ybar, fill=red!40!black, draw=none]  table [x expr=\coordindex, y={GPT-5}, col sep=comma] {tokens.csv};
          \addplot[ybar, fill=red!40!white, draw=none]  table [x expr=\coordindex, y={GPT-5 Instant}, col sep=comma] {tokens.csv};
          \addplot[ybar, fill=red!20!white, draw=none]  table [x expr=\coordindex, y={GPT-5 Thinking}, col sep=comma] {tokens.csv};
          \addplot[ybar, fill=brown!70!red, draw=none]  table [x expr=\coordindex, y={GPT-5 Pro}, col sep=comma] {tokens.csv};
          \addplot[ybar, fill=orange!80!red, draw=none] table [x expr=\coordindex, y={GPT-5 Auto}, col sep=comma] {tokens.csv};
          \addplot[ybar, fill=gray!50, draw=none]       table [x expr=\coordindex, y={Unknown}, col sep=comma] {tokens.csv};

          % --- FAMILY 2: Claude (Classic GREEN) ---
          % Pure "green" in LaTeX is extremely bright; "green!60!black" is the standard 'Forest Green' RGB look
          \addplot[ybar, fill=green!60!black, draw=none] table [x expr=\coordindex, y={Claude (All)}, col sep=comma] {tokens.csv};

          % --- FAMILY 3: Gemini (Classic BLUE) ---
          \addplot[ybar, fill=blue!80!black, draw=none]  table [x expr=\coordindex, y={Gemini-2.5-Pro}, col sep=comma] {tokens.csv};
          \addplot[ybar, fill=blue!50!white, draw=none]  table [x expr=\coordindex, y={Gemini-2.5-Flash}, col sep=comma] {tokens.csv};

          % --- AVERAGE LINE ---
          \draw [black, dashed] 
              ({rel axis cs:0,0}|-{axis cs:0,312696}) -- ({rel axis cs:1,0}|-{axis cs:0,312696});

          % --- LEGEND ---
          % Custom image for the Mean line
          \addlegendimage{line legend, black, dashed, very thick}
          
          % We only list the main models to keep the legend clean
          \legend{
              GPT-4o, GPT-o3, GPT-o4 Mini, GPT-5, GPT-5 Instant, GPT-5 Thinking, GPT-5 Pro, GPT-5 Auto, Unknown, 
              Claude (All), 
              Gemini Pro, Gemini Flash,
              {Avg: 312,696} 
          }

      \end{axis}
  \end{tikzpicture}
  \caption{\textbf{Distribution of LLM usage.} Bar chart displaying the total number of tokens generated by each participant in the LLM arm (n = 75), ordered by cumulative usage intensity. Two randomized participants (2/77) withdrew within 48 hours of study start without accessing the tools and are excluded from this visualization.}
  \label{fig:token_usage}
\end{figure}
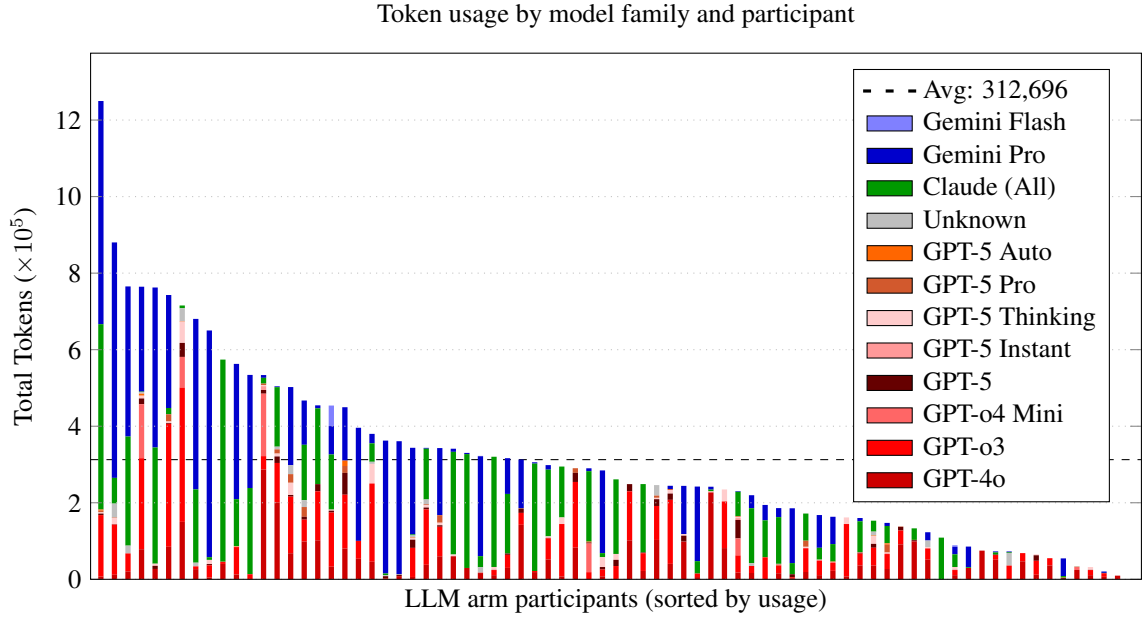

%% file: figures/ed-figure-1b.tex
\begin{figure}[H]
  \centering
  \begin{tikzpicture}[trim axis left]
  \begin{axis}[
    % --- Layout & Sizing (match your other figure) ---
    ybar stacked,
    width=\textwidth,
    height=6cm,
    scale only axis,
    trim axis left,
    % --- Axis Limits / Spacing ---
    ymin=0,
    enlarge x limits=0.12,
    bar width=22pt,
    % --- Y-Axis Formatting (fix ticks/units) ---
    scaled y ticks=false,
    ytick={0,2000000,4000000,6000000,8000000,10000000},
    yticklabels={0,2,4,6,8,10},
    ylabel={Total Tokens ($\times 10^6$)},
    % --- Grid & Style ---
    grid=major,
    grid style={dotted, gray!50},
    axis on top,
    % --- Title ---
    title={Token usage by model family (stacked within family)},
    % --- X Axis (add a hidden spacer category to push bars left) ---
    symbolic x coords={ChatGPT, Gemini, Claude, Spacer},
    xtick={ChatGPT, Gemini, Claude},
    xticklabel style={align=center},
    % --- Legend (upper-right, like your example) ---
    legend style={
      at={(0.97,0.97)},
      anchor=north east,
      fill=white!90,
      align=left,
      cells={anchor=west},
    },
    reverse legend,
  ]

  % -------- FAMILY 1: ChatGPT + Unknown (Classic RED) --------
  \addplot[ybar, fill=red!80!black, draw=none] coordinates {
    (ChatGPT, 3453125) (Gemini,0) (Claude,0) (Spacer,0)
  };
  \addlegendentry{GPT-4o}

  \addplot[ybar, fill=red, draw=none] coordinates {
    (ChatGPT, 4503510) (Gemini,0) (Claude,0) (Spacer,0)
  };
  \addlegendentry{GPT-o3}

  \addplot[ybar, fill=red!60!white, draw=none] coordinates {
    (ChatGPT, 512869) (Gemini,0) (Claude,0) (Spacer,0)
  };
  \addlegendentry{GPT-o4 Mini}

  \addplot[ybar, fill=red!40!black, draw=none] coordinates {
    (ChatGPT, 466301) (Gemini,0) (Claude,0) (Spacer,0)
  };
  \addlegendentry{GPT-5}

  \addplot[ybar, fill=red!20!white, draw=none] coordinates {
    (ChatGPT, 413275) (Gemini,0) (Claude,0) (Spacer,0)
  };
  \addlegendentry{GPT-5 Thinking}

  \addplot[ybar, fill=brown!70!red, draw=none] coordinates {
    (ChatGPT, 190965) (Gemini,0) (Claude,0) (Spacer,0)
  };
  \addlegendentry{GPT-5 Pro}

  \addplot[ybar, fill=orange!80!red, draw=none] coordinates {
    (ChatGPT, 38294) (Gemini,0) (Claude,0) (Spacer,0)
  };
  \addlegendentry{GPT-5 Auto}

  \addplot[ybar, fill=red!40!white, draw=none] coordinates {
    (ChatGPT, 17291) (Gemini,0) (Claude,0) (Spacer,0)
  };
  \addlegendentry{GPT-5 Instant}

  \addplot[ybar, fill=gray!50, draw=none] coordinates {
    (ChatGPT, 294338) (Gemini,0) (Claude,0) (Spacer,0)
  };
  \addlegendentry{Unknown}

  % -------- FAMILY 2: Claude (Classic GREEN) --------
  \addplot[ybar, fill=green!60!black, draw=none] coordinates {
    (ChatGPT,0) (Gemini,0) (Claude,6260236) (Spacer,0)
  };
  \addlegendentry{Claude}

  % -------- FAMILY 3: Gemini (Classic BLUE) --------
  \addplot[ybar, fill=blue!80!black, draw=none] coordinates {
    (ChatGPT,0) (Gemini,7868564) (Claude,0) (Spacer,0)
  };
  \addlegendentry{Gemini 2.5 Pro}

  \addplot[ybar, fill=blue!50!white, draw=none] coordinates {
    (ChatGPT,0) (Gemini,58797) (Claude,0) (Spacer,0)
  };
  \addlegendentry{Gemini 2.5 Flash}

  \end{axis}
  \end{tikzpicture}
  \caption{\textbf{Total Token Usage by LLM Family.}Stacked segments represent the proportion of tokens generated by each model family: OpenAI (red), Anthropic (green), and Google DeepMind (blue). The distribution highlights substantial usage of advanced frontier models, with Gemini 2.5 (32.6\%), Claude (25.7\%), and OpenAI models (41.7\% combined) accounting for the total volume. GPT-5 usage was introduced on week 5 of the study.}
  \label{fig:tokens_by_family_stacked}
\end{figure}
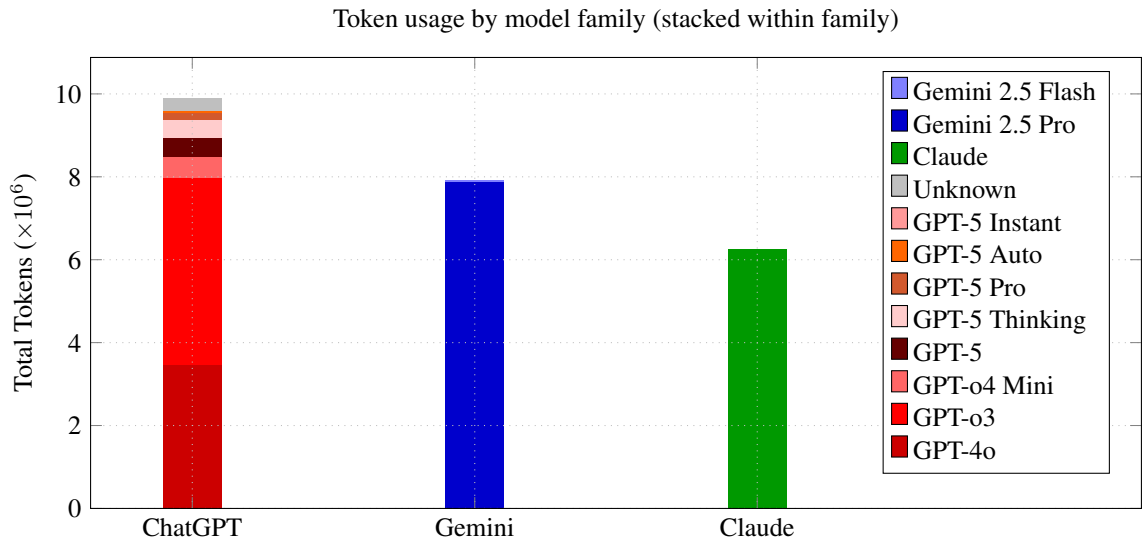

%% file: figures/ed-figure-2.tex
\begin{figure}[h]
  \centering
    \includegraphics[
        page=1,
        width=\textwidth,
        angle=0
    ]{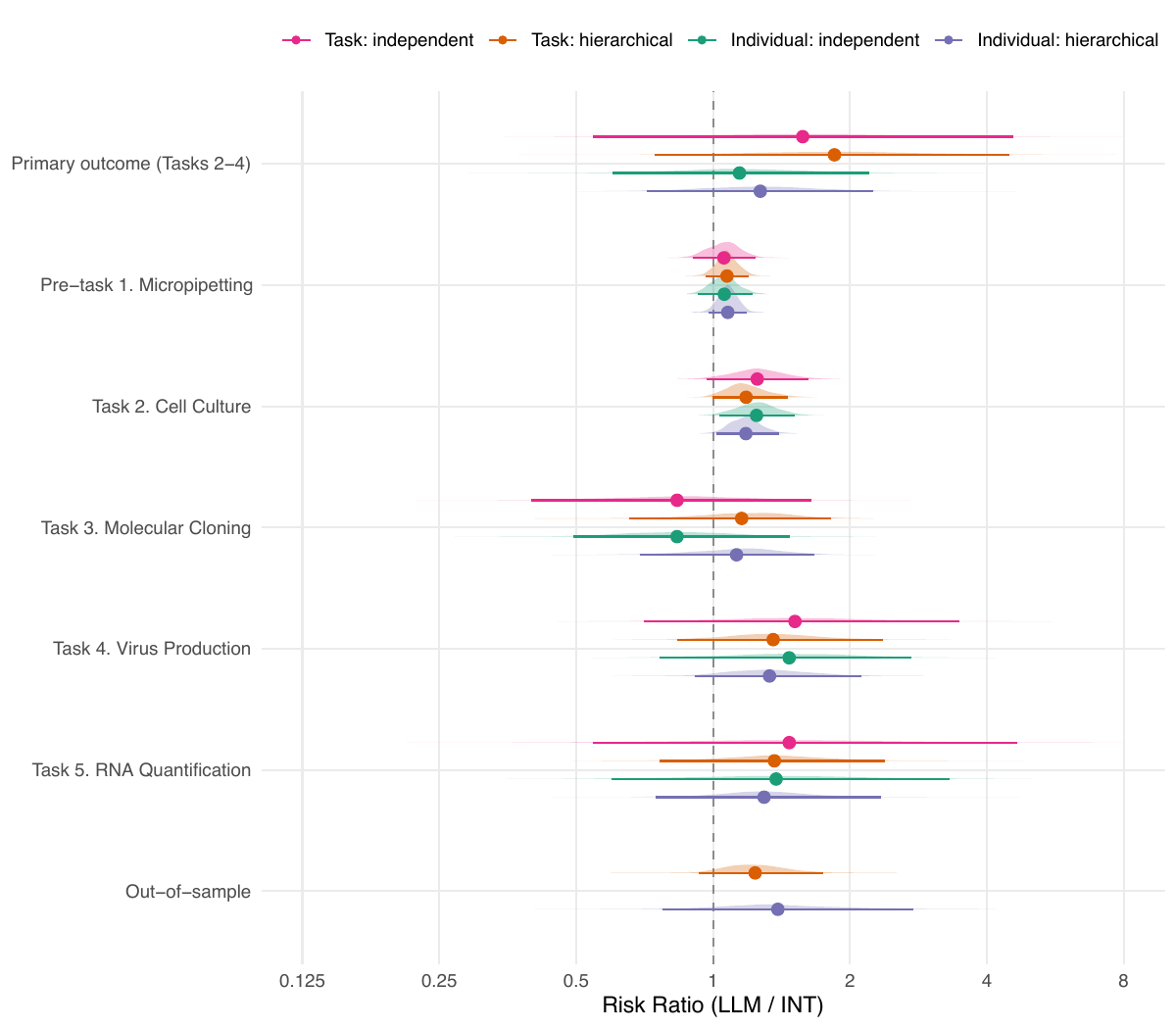}
  \caption{\textbf{Sensitivity analysis of Bayesian model specifications.} Posterior distributions of the Risk Ratio (RR) comparing an independent task-level model (pink), a hierarchical task-level model (orange), an independent participant-level model (green), and the primary hierarchical participant-level model (purple). Colored markers represent posterior estimates from a hierarchical Bayesian logistic regression model, displaying posterior means, 95\% credible intervals (CrI), and the posterior probability of a positive effect (\(\Pr(RR) > 1\)). Shaded regions depict full posterior densities.}
  \label{fig:ed-figure-2}
\end{figure}

%% file: figures/ed-figure-3.tex
\begin{figure}[htbp]
  \centering

  % Each panel is ~half the line width, so that two fit per row
  \begin{subfigure}[t]{0.49\textwidth}
    \centering
    \includegraphics[width=\linewidth]{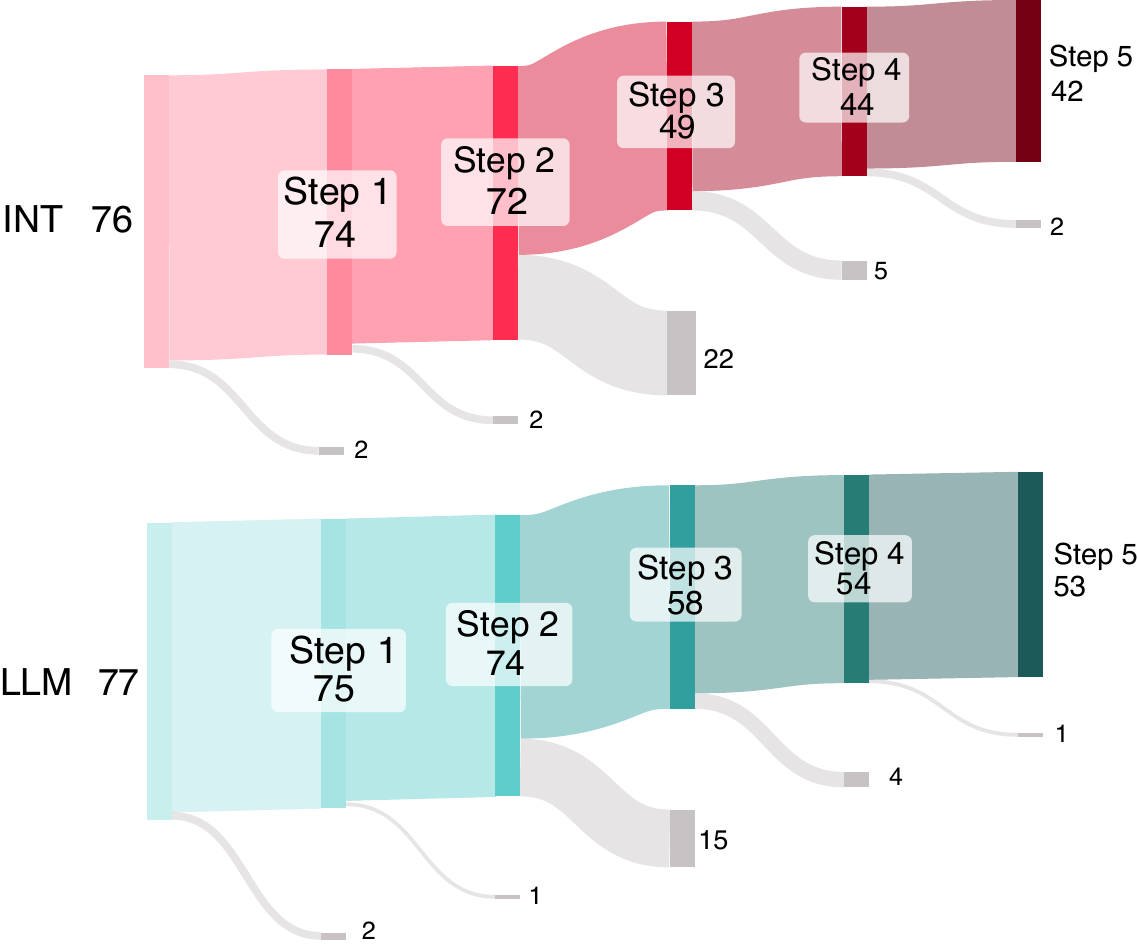}
    \caption{Task 2: Cell Culture}
  \end{subfigure}\hfill
  \begin{subfigure}[t]{0.49\textwidth}
    \centering
    \includegraphics[width=\linewidth]{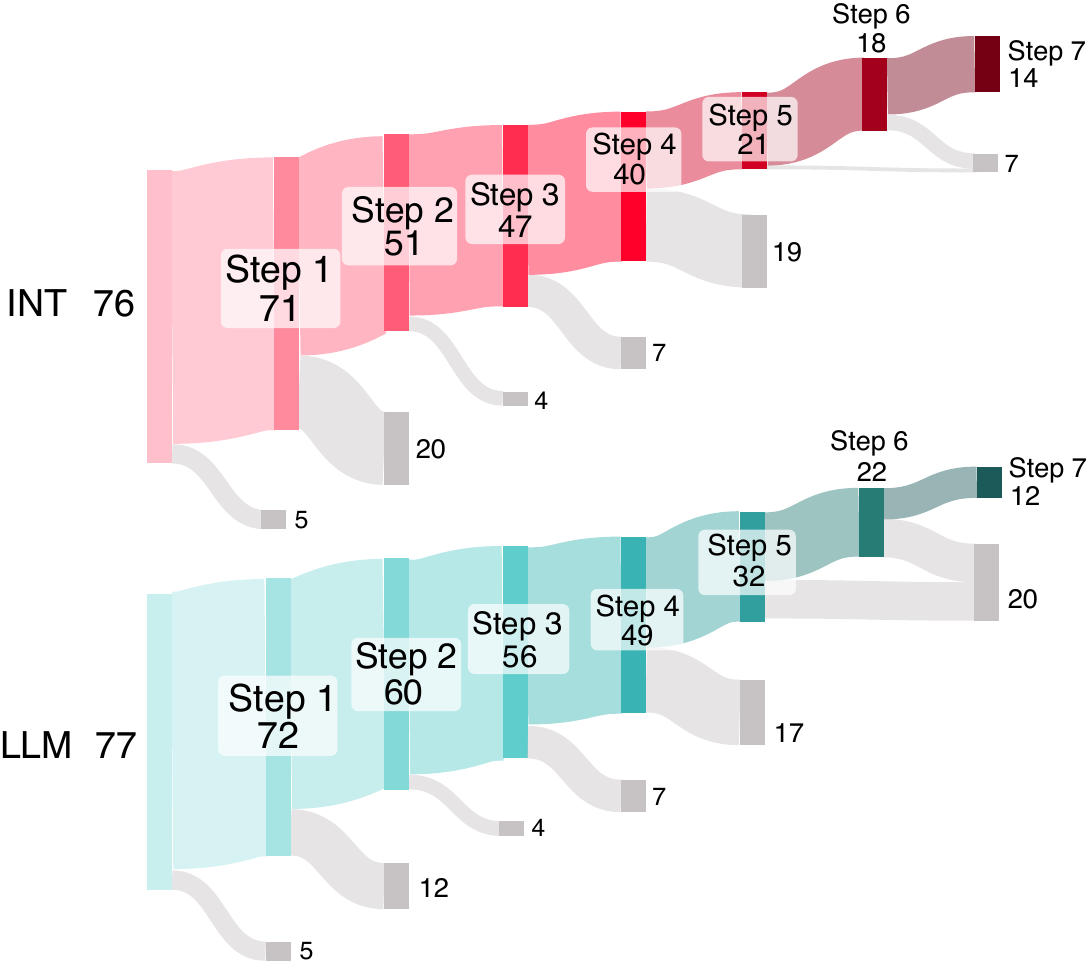}
    \caption{Task 3: Molecular Cloning}
  \end{subfigure}

  \vspace{0.5em}

  \begin{subfigure}[t]{0.49\textwidth}
    \centering
    \includegraphics[width=\linewidth]{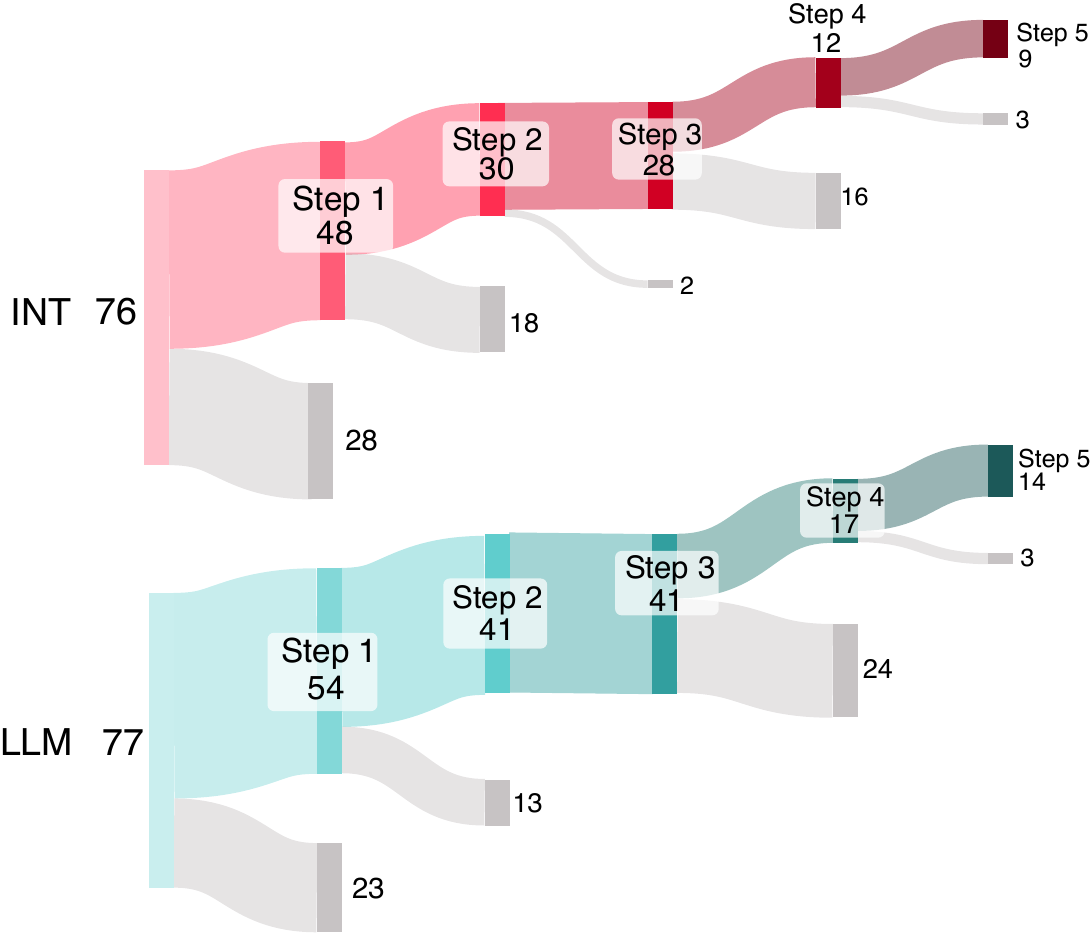}
    \caption{Task 4: Virus Production}
  \end{subfigure}\hfill
  \begin{subfigure}[t]{0.49\textwidth}
    \centering
    \includegraphics[width=\linewidth]{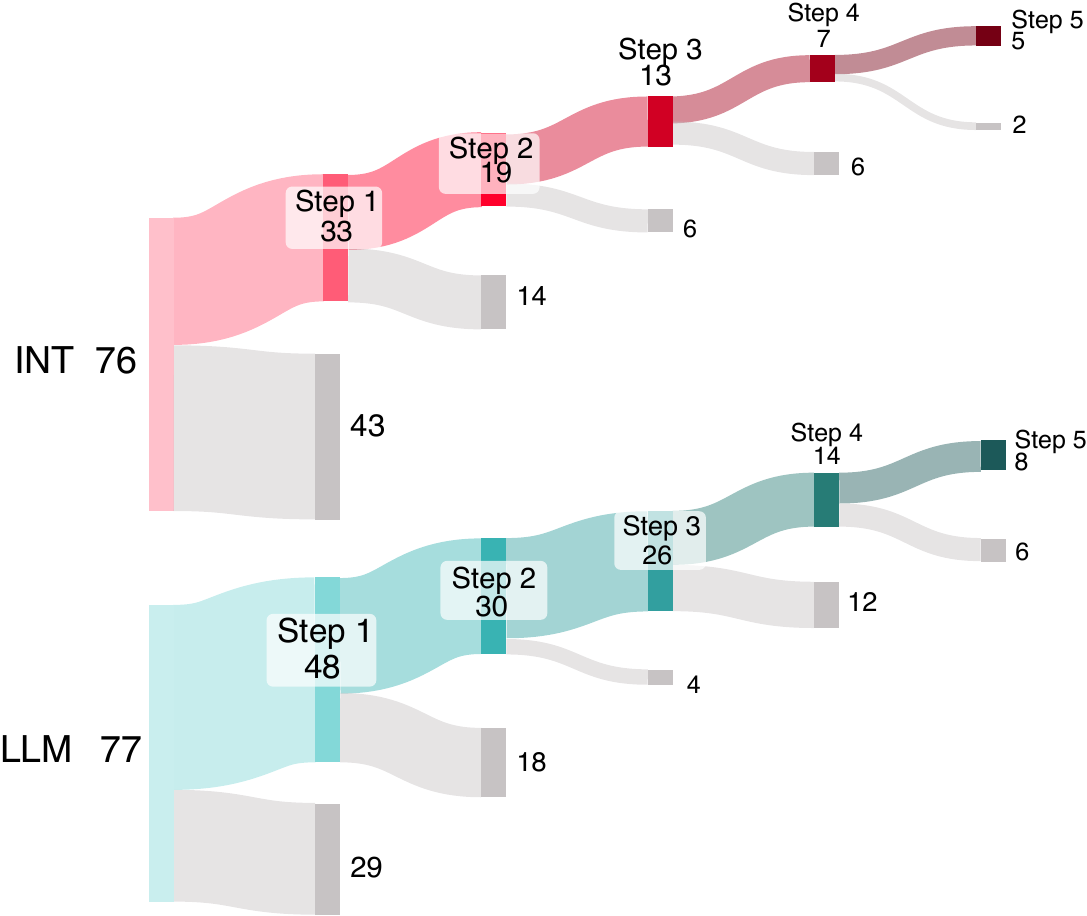}
    \caption{Task 5: RNA Quantification}
  \end{subfigure}
  \caption{\textbf{Participant progression through procedural steps of each task. A-D.} Sankey diagrams illustrating participant flow through sequential milestones for Task 2 (a), Task 3 (b), Task 4 (c), and Task 5 (d). Ribbons represent the Internet arm (pink, n = 76) and LLM arm (blue, n = 77), with stream width proportional to the number of participants successfully completing each stage. Gray paths diverging downwards denote attrition (failure to complete the step or discontinuation of the task). Numbers within nodes indicate the count of participants who succeeded at each step. Detailed definitions of specific procedural steps are provided in \autoref{tab:ed-table-15}.}
  \label{fig:ed-figure-3}
\end{figure}

%% file: figures/ed-figure-4.tex
\begin{figure}[h]
  \centering
    \includegraphics[
        page=1,
        width=\textwidth,
        angle=0
    ]{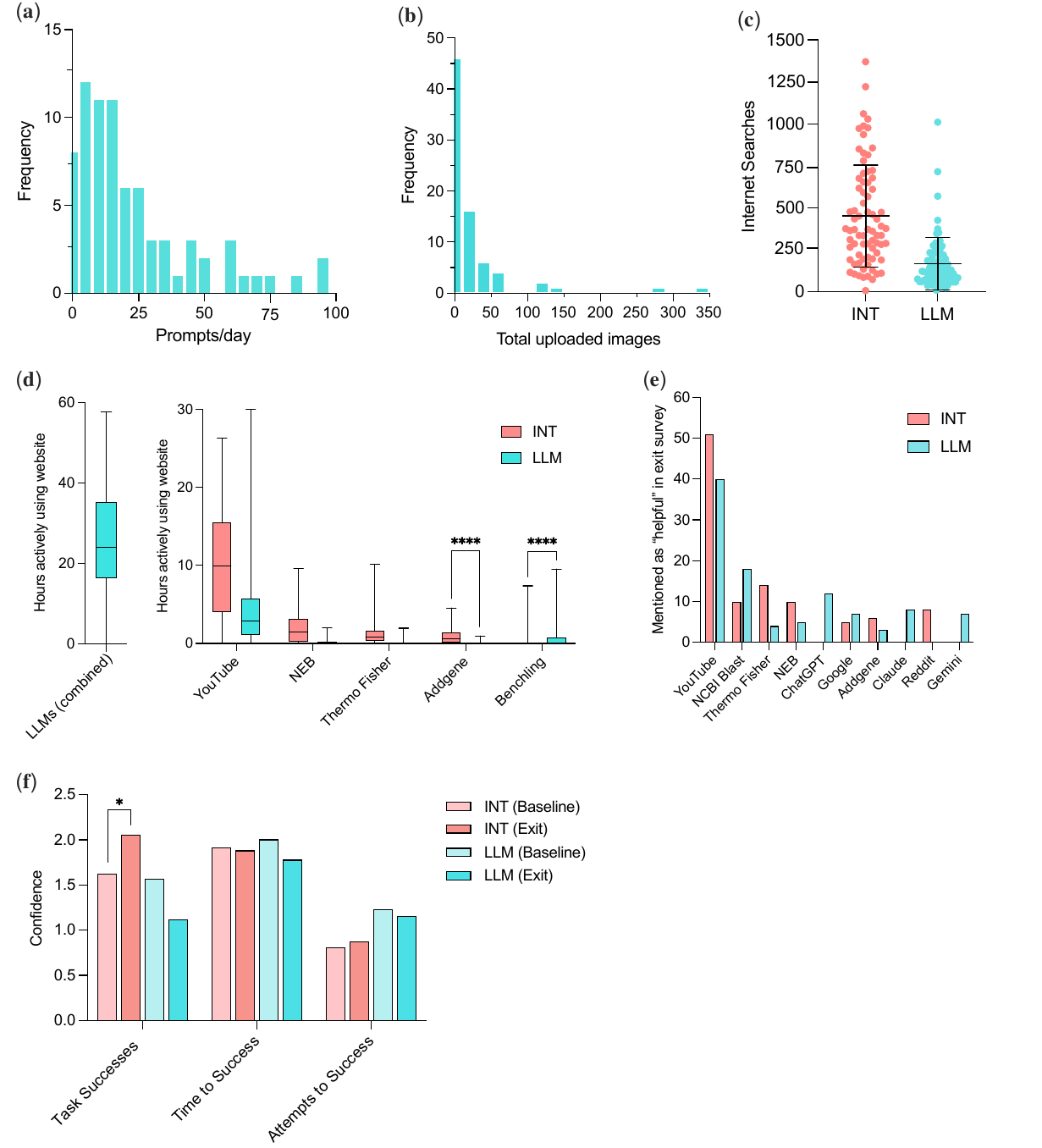}
  \caption{\textbf{LLM usage patterns and participant perceptions.} \textbf{a}, Frequency distribution of daily LLM prompt volume per participant (\(n = 77\) LLM arm). Prompt counts were derived from automated scripts quantifying \enquote{user} messages. \textbf{b}, Frequency distribution of cumulative image uploads to LLM interfaces. Note: Analysis is restricted to OpenAI and Gemini platforms as Claude data lacked file export capabilities. \textbf{c}, Comparison of total internet search queries. Participants in the Internet arm executed a higher volume of traditional searches compared to the LLM arm. textbf{d}, Active engagement time with digital resources. Box plots of total hours spent interacting with LLM interfaces (left) and most commonly used websites (right) (**\(P < 0.001\), two-way ANOVA with multiple comparisons). textbf{e}, Resources most frequently cited as \enquote{helpful} in exit surveys. \textbf{f}, Longitudinal assessment of self-predicted performance. Bar chart of participants' estimates of confidence in succeeding at tasks, speed (\enquote{Time to Success}), and efficiency (\enquote{Attempts to Success}) at week 2 (Baseline) and week 8 (study exit). Data represent mean \(\pm\) s.d. (\(P < 0.05\), two-way ANOVA with Šídák's multiple comparisons test)}
  \label{fig:ed-figure-4}
\end{figure}

%% file: figures/ed-figure-5.tex
\begin{figure}[h]
  \centering
    % Left panel
    \begin{subfigure}[t]{0.49\textwidth}
      \centering
      \includegraphics[width=\linewidth]{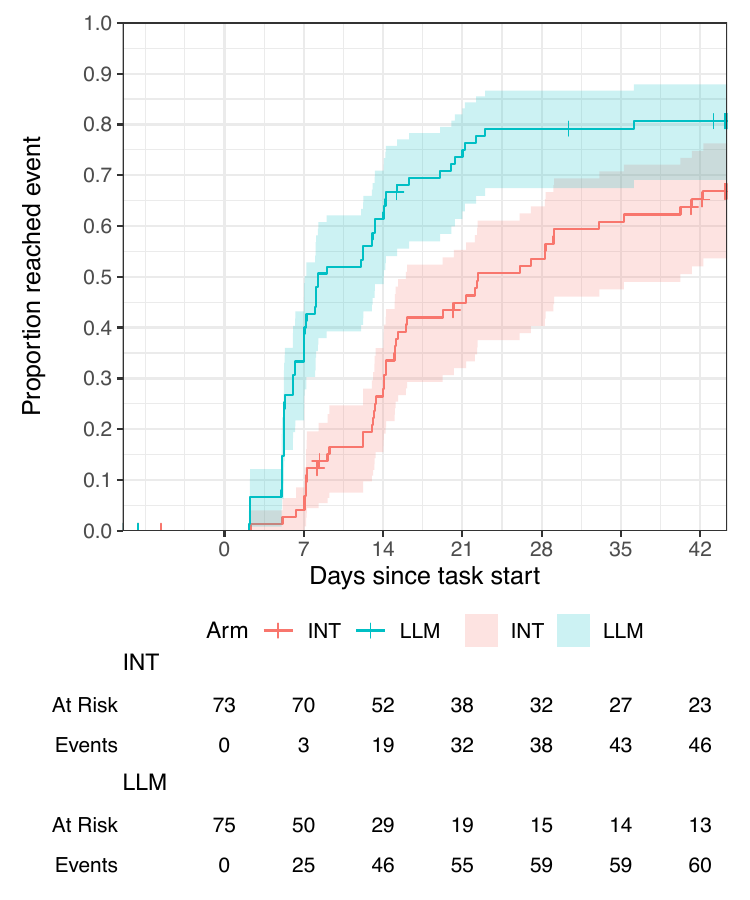}
      \caption{Time to First Request}
      \label{fig:fig-5a}
    \end{subfigure}\hfill
    % Right panel
    \begin{subfigure}[t]{0.49\textwidth}
      \centering
      \includegraphics[width=\linewidth]{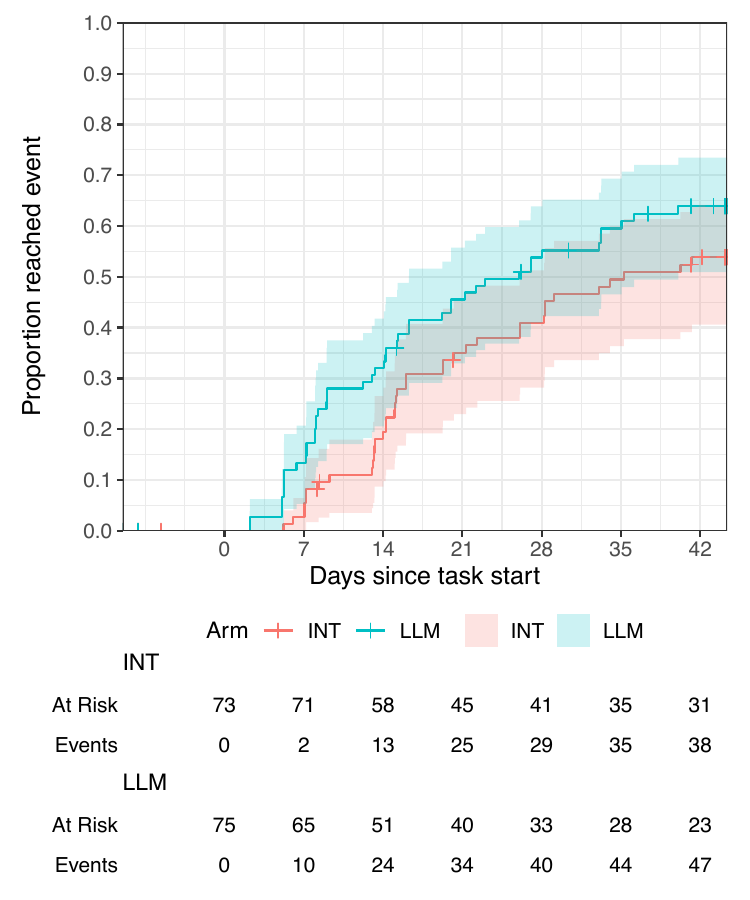}
      \caption{Time to First Correct Request}
      \label{fig:fig-5b}
    \end{subfigure}
  \caption{\textbf{Time to correct reagent submission for molecular cloning (Task 3).} \textbf{(a)} Kaplan-Meier cumulative incidence curves showing the time to first reagent request submission for the Internet arm (pink) and LLM arm (blue). Shaded regions represent 95\% confidence intervals (CI). \textbf{(b)}, Kaplan-Meier cumulative incidence curves showing the time to first correct reagent request submission. Shaded regions represent 95\% confidence intervals (CI)}
  \label{fig:fig-5}
\end{figure}

%% file: sections/10-extended-data-tables.tex
\section{Extended Data Tables}
% We ensure that all tables in this section are prefixed with "Extended Data Table \n:"
% \setcounter{table}{0}
% \renewcommand{\tablename}{Extended Data Table}

\input{tables/ed-table-1.tex}
\input{tables/ed-table-2.tex}
\input{tables/ed-table-3.tex}
\clearpage
\input{tables/ed-table-4.tex}
\input{tables/ed-table-5.tex}
\input{tables/ed-table-6.tex}
\input{tables/ed-table-7.tex}
\input{tables/ed-table-8.tex}
\input{tables/ed-table-9.tex}
\input{tables/ed-table-10.tex}
\input{tables/ed-table-11.tex}
\input{tables/ed-table-12.tex}
\input{tables/ed-table-13.tex}
\input{tables/ed-table-14.tex}
\input{tables/ed-table-15.tex}
\clearpage
\input{tables/ed-table-16.tex}
\input{tables/ed-table-17.tex}
\input{tables/ed-table-18.tex}

%% file: tables/ed-table-1.tex
\begin{table}[H]
    \caption{Definition of Access to the Internet}
    \label{tab:ed-table-1}
    \centering

    % @{} removes the default padding on the far left and right for a flush look
    \begin{tabularx}{\textwidth}{@{} p{5.5cm} l X @{}} 
        \toprule

        \textbf{Internet Resource} & \textbf{Inclusion} & \textbf{Rationale for} \\ 
        (Examples) & & \textbf{Inclusion/Exclusion} \\ \midrule

        % --- GROUP 1: YES ---
        \textbf{Websites} \newline (Wikipedia, YouTube, ThermoFisher) & Yes & \multirow{13}{=}{These resources represent standard and commonly accessible sources of online biology and laboratory knowledge, similar to what novices may access in a real-world setting. Image search engines may be used by participants to perform reverse lookups of unfamiliar reagents or equipment. Participants may utilize preprint servers, open-access databases, and open-access textbooks.} \\ \cmidrule(r){1-2}

        \textbf{Search engines} \newline (Google, Bing, Duckduckgo) & Yes & \\ \cmidrule(r){1-2}

        \textbf{Image search engines} \newline (Google Images, TinEye) & Yes & \\ \cmidrule(r){1-2}

        \textbf{Public Databases} \newline (PubMed, Arxiv, Bioarxiv) & Yes & \\ \cmidrule(r){1-2}

        \textbf{Educational Platforms} \newline (Coursera, Khanacademy Biology) & Yes & \\ \cmidrule(r){1-2}

        \textbf{Online Textbooks} \newline (OpenStax Biology, Wikibooks) & Yes & \\ \midrule

        % --- GROUP 2: READ ONLY ---
        \textbf{Q\&A Sites} \newline (Biology StackExchange, Quora) & Read Only & \multirow{5}{=}{Participants may access publicly posted content but may not interact in any way (e.g., posting, commenting, or voting). This restriction prevents real-time external assistance.} \\ \cmidrule(r){1-2}

        \textbf{Discussion Forums} \newline (Reddit, ResearchGate Forums) & Read Only & \\ \cmidrule(r){1-2}

        \textbf{Social Media} \newline (Facebook Groups, Twitter) & Read Only & \\ \midrule

        % --- GROUP 3: NO ---
        \textbf{Online Chatrooms} \newline (Biology Discord, IRC) & No & \multirow{6}{=}{These platforms enable real-time communication and could introduce external assistance. Participants will be instructed to not access these resources, and the resources will be blocked via technical measures.} \\ \cmidrule(r){1-2}

        \textbf{Messaging Apps} \newline (Facebook Messenger, WhatsApp) & No & \\ \cmidrule(r){1-2}

        \textbf{Email} \newline (Sending an email for advice) & No & \\ \bottomrule

    \end{tabularx}
\end{table}

%% file: tables/ed-table-2.tex
\begin{table}[H]
    \caption{Definition of Access to LLMs}
    \label{tab:ed-table-2}
    \centering
    \begin{tabularx}{\textwidth}{@{} p{5.5cm} l X @{}}
        \toprule
        \textbf{LLM-Based Resource} & \textbf{Inclusion} & \textbf{Rationale for} \\
        (Examples) & & \textbf{Inclusion/Exclusion} \\ \midrule

        % --- GROUP 1: YES (Merged Rationale) ---
        \textbf{ChatGPT Family of Models}, by OpenAI, inclusive but not limited to ChatGPT o3. & Yes & \multirow{9}{=}{The three top-performing models on biology benchmarks such as ProtocolQA, Virology Capabilities Test (VCT) were selected for inclusion into the study. By using LLMs with proven capabilities in biology, the study standardizes the quality of AI assistance provided to participants} \\ \cmidrule(r){1-2}

        \textbf{Claude Family of Models}, by Anthropic, inclusive but not limited to Claude 4 Opus. & Yes & \\ \cmidrule(r){1-2}

        \textbf{Gemini Family of Models}, by Google DeepMind, inclusive but not limited to Gemini 2.5 Pro. & Yes & \\ \midrule

        % --- GROUP 2: NO ---
        \textbf{Other LLMs outside the above specification} \newline (e.g., Grok, Github Copilot, Reddit Answers) & No & In order to reduce variations in the strength of treatment offered, participants will be blocked from accessing LLMs outside of the above specification. \\ \bottomrule

    \end{tabularx}
\end{table}

%% file: tables/ed-table-3.tex
\begin{table}[h]
    \caption{Enrollment and Randomization}
    \label{tab:ed-table-3}
    \centering
    % tabular* with @{\extracolsep{\fill}} automatically spaces columns evenly
    \begin{tabular*}{\textwidth}{@{\extracolsep{\fill}} lcccccc @{}}
        \toprule
        & \multicolumn{3}{c}{\textbf{Full Analysis Set (FAS)}} & \multicolumn{3}{c}{\textbf{Per Protocol Set (PPS)}} \\
        \cmidrule(lr){2-4} \cmidrule(l){5-7}
        & \textbf{Internet} & \textbf{LLM} & \textbf{Total} & \textbf{Internet} & \textbf{LLM} & \textbf{Total} \\ 
        \midrule
        Randomized & 76 & 77 & 153 & 64 & 64 & 128 \\ 
        \bottomrule
    \end{tabular*}
\end{table}

%% file: tables/ed-table-4.tex
\begin{table}[H]
    \caption{Demographics \& Background}
    \label{tab:ed-table-4}
    \centering
    % X column allows the long labels to wrap text if needed
    % l column for the stats ensures the equations align neatly to the left
    \begin{tabularx}{\textwidth}{@{} X c c l @{}} 
        \toprule
        \textbf{A. Full Analysis Set (FAS)} & \textbf{Internet} & \textbf{LLM} & \textbf{Balance Test} \\ 
        \midrule

        % --- AGE SECTION ---
        \textit{Age (years)} & 24.0 (11.5) & 23.1 (8.6) & $t(133.81) = 0.52, P = 0.605, d = 0.085$ \\ 
        \addlinespace

        % --- SEX SECTION ---
        \textbf{Sex assigned at birth} & & & \\
        \hspace{3mm} Female & 42 (55.3\%) & 41 (53.2\%) & $P = 0.837, V = 0.047$ \\
        \hspace{3mm} Male & 31 (40.8\%) & 34 (44.2\%) & \\
        \hspace{3mm} Other & 0 (0.0\%) & 0 (0.0\%) & \\
        \hspace{3mm} Missing & 3 (3.9\%) & 2 (2.6\%) & \\ 
        \addlinespace

        % --- RACE SECTION ---
        \textbf{Race/ethnicity} & & & \\
        \hspace{3mm} Hispanic or Latino & 7 (9.2\%) & 11 (14.3\%) & $P = 0.453, V = 0.058$ \\
        \hspace{3mm} American Indian or & 1 (1.3\%) & 1 (1.3\%) & $P = 1.00, V = 0.000$ \\
        \hspace{3mm} Alaska Native & & & \\ % Manual break for clarity if needed, or let X wrap it
        \hspace{3mm} Asian & 25 (32.9\%) & 26 (33.8\%) & $P = 1.00, V = 0.000$ \\
        \hspace{3mm} Black or African & 11 (14.5\%) & 13 (16.9\%) & $P = 0.825, V = 0.015$ \\
        \hspace{3mm} American & & & \\
        \hspace{3mm} Middle Eastern or & 4 (5.3\%) & 1 (1.3\%) & $P = 0.209, V = 0.075$ \\
        \hspace{3mm} North African & & & \\
        \hspace{3mm} Native Hawaiian or & 1 (1.3\%) & 0 (0.0\%) & $P = 0.497, V = 0.001$ \\
        \hspace{3mm} Pacific Islander & & & \\
        \hspace{3mm} White & 31 (40.8\%) & 26 (33.8\%) & $P = 0.406, V = 0.059$ \\
        \hspace{3mm} Multiple selections & 10 (13.2\%) & 6 (7.8\%) & $P = 0.304, V = 0.066$ \\
        \hspace{3mm} (derived) & & & \\
        \hspace{3mm} Prefer not to answer & 3 (3.9\%) & 3 (3.9\%) & $P = 1.000, V = 0.000$ \\
        \hspace{3mm} Missing & 4 (5.3\%) & 2 (2.6\%) & $P = 0.442, V = 0.035$ \\ 
        \bottomrule
    \end{tabularx}

    \vspace{1em} % Adds a little breathing room between the two tables

    % Using the same column setup: X for labels, c for data, l for stats
    \begin{tabularx}{\textwidth}{@{} X c c l @{}} 
        \toprule
        \textbf{B. Per Protocol Set (PPS)} & \textbf{Internet} & \textbf{LLM} & \textbf{Balance Test} \\ 
        \midrule

        % --- AGE SECTION ---
        \textit{Age (years)} & 24.0 (11.9) & 23.1 (9.0) & $t(117.34) = 0.51, P = 0.610, d = 0.090$ \\ 
        \addlinespace

        % --- SEX SECTION ---
        \textbf{Sex assigned at birth} & & & \\
        \hspace{3mm} Female & 35 (54.7\%) & 36 (56.2\%) & $P = 1.000, V = 0.000$ \\
        \hspace{3mm} Male & 29 (45.3\%) & 28 (43.8\%) & \\
        \hspace{3mm} Other & 0 (0.0\%) & 0 (0.0\%) & \\
        \hspace{3mm} Missing & 0 (0.0\%) & 0 (0.0\%) & \\ 
        \addlinespace

        % --- RACE SECTION ---
        \textbf{Race/ethnicity} & & & \\
        \hspace{3mm} Hispanic or Latino & 5 (7.8\%) & 10 (15.6\%) & $P = 0.271, V = 0.097$ \\
        \hspace{3mm} American Indian or & 0 (0.0\%) & 1 (1.6\%) & $P = 1.000, V = 0.000$ \\
        \hspace{3mm} Alaska Native & & & \\
        \hspace{3mm} Asian & 22 (34.4\%) & 23 (35.9\%) & $P = 1.000, V = 0.000$ \\
        \hspace{3mm} Black or African & 9 (14.1\%) & 9 (14.1\%) & $P = 1.000, V = 0.000$ \\
        \hspace{3mm} American & & & \\
        \hspace{3mm} Middle Eastern or & 3 (4.7\%) & 1 (1.6\%) & $P = 0.619, V = 0.045$ \\
        \hspace{3mm} North African & & & \\
        \hspace{3mm} Native Hawaiian or & 1 (1.6\%) & 0 (0.0\%) & $P = 1.000, V = 0.000$ \\
        \hspace{3mm} Pacific Islander & & & \\
        \hspace{3mm} White & 29 (45.3\%) & 23 (35.9\%) & $P = 0.368, V = 0.080$ \\
        \hspace{3mm} Multiple selections & 8 (12.5\%) & 5 (7.8\%) & $P = 0.560, V = 0.052$ \\
        \hspace{3mm} (derived) & & & \\
        \hspace{3mm} Prefer not to answer & 3 (4.7\%) & 2 (3.1\%) & $P = 1.000, V = 0.000$ \\
        \hspace{3mm} Missing & 1 (1.6\%) & 0 (0.0\%) & $P = 1.000, V = 0.000$ \\ 
        \bottomrule
    \end{tabularx}
\end{table}

\emph{Note: Balance tests were conducted row wise using Welch's t-tests (for continuous variables) and Fisher's exact tests (for categorical variables; mutually exclusive categories were grouped together into one analysis). Effect size was quantified using Cohen's \(d\) (for continuous variables) and Cramer's \(V\) (for categorical variables).}

%% file: tables/ed-table-5.tex
\begin{table}[h]
    \caption{Baseline Educational Characteristics}
    \label{tab:ed-table-5}
    \centering
    \begin{tabularx}{\textwidth}{@{} X c c l @{}} 
        \toprule
        \textbf{A. Full Analysis Set (FAS)} & \textbf{Internet} & \textbf{LLM} & \textbf{Balance Test} \\ 
        \midrule

        % --- EDUCATIONAL ATTAINMENT ---
        \textbf{\textit{Educational attainment}} & & & \\
        \hspace{3mm} Less than high school & 0 (0.0\%) & 0 (0.0\%) & $P = 0.256, V = 0.231$ \\
        \hspace{3mm} High school diploma or GED & 11 (14.5\%) & 11 (14.3\%) & \\
        \hspace{3mm} Some college ($<1$ year) & 0 (0.0\%) & 1 (1.3\%) & \\
        \hspace{3mm} Some college ($\ge 1$ year), no degree & 49 (64.5\%) & 45 (58.4\%) & \\
        \hspace{3mm} Associate degree & 5 (6.6\%) & 2 (2.6\%) & \\
        \hspace{3mm} Bachelor's degree & 3 (3.9\%) & 12 (15.6\%) & \\
        \hspace{3mm} Master's degree & 4 (5.3\%) & 3 (3.9\%) & \\
        \hspace{3mm} Professional degree (e.g., MD, JD) & 0 (0.0\%) & 0 (0.0\%) & \\
        \hspace{3mm} Doctorate (e.g., PhD, EdD) & 0 (0.0\%) & 0 (0.0\%) & \\
        \hspace{3mm} Prefer not to answer & 1 (1.3\%) & 1 (1.3\%) & \\
        \hspace{3mm} Missing & 3 (3.9\%) & 2 (2.6\%) & \\
        \addlinespace

        % --- FIELD OF STUDY ---
        \textbf{\textit{Field of study}} & & & \\
        \hspace{3mm} STEM & 67 (88.2\%) & 67 (87.0\%) & $P = 0.744, V = 0.117$ \\
        \hspace{3mm} Non-STEM & 5 (6.6\%) & 4 (5.2\%) & \\
        \hspace{3mm} Interdisciplinary/Other & 1 (1.3\%) & 0 (0.0\%) & \\
        \hspace{3mm} Missing & 3 (3.9\%) & 6 (7.8\%) & \\
        \bottomrule
    \end{tabularx}

    \vspace{1em} % Adds a little breathing room between the two tables

    \begin{tabularx}{\textwidth}{@{} X c c l @{}} 
        \toprule
        \textbf{B. Per Protocol Set (PPS)} & \textbf{Internet} & \textbf{LLM} & \textbf{Balance Test} \\ 
        \midrule

        % --- EDUCATIONAL ATTAINMENT ---
        \textbf{\textit{Educational attainment}} & & & \\
        \hspace{3mm} Less than high school & 0 (0.0\%) & 0 (0.0\%) & $P = 0.405, V = 0.213$ \\
        \hspace{3mm} High school diploma or GED & 9 (14.1\%) & 8 (12.5\%) & \\
        \hspace{3mm} Some college ($<1$ year) & 0 (0.0\%) & 1 (1.6\%) & \\
        \hspace{3mm} Some college ($\ge 1$ year), no degree & 43 (67.2\%) & 39 (60.9\%) & \\
        \hspace{3mm} Associate degree & 4 (6.2\%) & 2 (3.1\%) & \\
        \hspace{3mm} Bachelor's degree & 3 (4.7\%) & 10 (15.6\%) & \\
        \hspace{3mm} Master's degree & 4 (6.2\%) & 3 (4.7\%) & \\
        \hspace{3mm} Professional degree (e.g., MD, JD) & 0 (0.0\%) & 0 (0.0\%) & \\
        \hspace{3mm} Doctorate (e.g., PhD, EdD) & 0 (0.0\%) & 0 (0.0\%) & \\
        \hspace{3mm} Prefer not to answer & 1 (1.6\%) & 1 (1.6\%) & \\
        \hspace{3mm} Missing & 0 (0.0\%) & 0 (0.0\%) & \\
        \addlinespace

        % --- FIELD OF STUDY ---
        \textbf{\textit{Field of study}} & & & \\
        \hspace{3mm} STEM & 59 (92.2\%) & 56 (87.5\%) & $P = 0.173, V = 0.199$ \\
        \hspace{3mm} Non-STEM & 4 (6.2\%) & 4 (6.2\%) & \\
        \hspace{3mm} Interdisciplinary/Other & 1 (1.6\%) & 0 (0.0\%) & \\
        \hspace{3mm} Missing & 0 (0.0\%) & 4 (6.2\%) & \\
        \bottomrule
    \end{tabularx}
\end{table}

\emph{Note: Balance tests were conducted using Fisher's exact tests (mutually exclusive categories were grouped together into one analysis). Effect size was quantified using Cramer's \(V\).}

%% file: tables/ed-table-6.tex
\begin{table}[H]
    \caption{Prior Biology Experience}
    \label{tab:ed-table-6}
    \centering

    % --- TABLE A: FULL ANALYSIS SET ---
    \begin{tabularx}{\textwidth}{@{} X c c l @{}} 
        \toprule
        \textbf{A. Full Analysis Set (FAS)} & \textbf{Internet} & \textbf{LLM} & \textbf{Balance Test} \\ 
        \midrule

        % --- BIO-EXP TOTAL ---
        \textbf{BIO-EXP total (0--30)} & & & \\
        \hspace{3mm} Mean (SD) & 4.4 (4.7) & 5.0 (4.6) & $t(145.45) = -0.68, P = 0.497, d = -0.112$ \\
        \hspace{3mm} Median (IQR) & 3.0 (7.0) & 4.0 (6.5) & \\
        \addlinespace

        % --- PIPETTING ---
        \textbf{Pipetting subscore (0--6)} & & & \\
        \hspace{3mm} Mean (SD) & 2.3 (2.2) & 2.6 (2.1) & $t(145.55) = -0.73, P = 0.467, d = -0.120$ \\
        \hspace{3mm} Median (IQR) & 2.0 (4.0) & 2.0 (4.0) & \\
        \addlinespace

        % --- CELL CULTURE ---
        \textbf{Cell culture subscore (0--6)} & & & \\
        \hspace{3mm} Mean (SD) & 0.5 (0.9) & 0.6 (1.1) & $t(142.48) = -0.62, P = 0.536, d = -0.102$ \\
        \hspace{3mm} Median (IQR) & 0.0 (1.0) & 0.0 (1.0) & \\
        \addlinespace

        % --- MOLECULAR CLONING ---
        \textbf{Molecular cloning subscore (0--6)} & & & \\
        \hspace{3mm} Mean (SD) & 0.6 (1.3) & 0.6 (1.2) & $t(145.14) = -0.25, P = 0.802, d = -0.041$ \\
        \hspace{3mm} Median (IQR) & 0.0 (0.0) & 0.0 (1.0) & \\
        \addlinespace

        % --- TRANSFECTION ---
        \textbf{Transfection / AAV subscore (0--6)} & & & \\
        \hspace{3mm} Mean (SD) & 0.2 (0.6) & 0.1 (0.3) & $t(108.77) = 0.97, P = 0.333, d = 0.161$ \\
        \hspace{3mm} Median (IQR) & 0.0 (0.0) & 0.0 (0.0) & \\
        \addlinespace

        % --- qPCR ---
        \textbf{qPCR subscore (0--6)} & & & \\
        \hspace{3mm} Mean (SD) & 0.9 (1.1) & 1.1 (1.3) & $t(143.17) = -0.90, P = 0.370, d = -0.147$ \\
        \hspace{3mm} Median (IQR) & 1.0 (1.0) & 1.0 (2.0) & \\
        \bottomrule
    \end{tabularx}

    \vspace{1em} % Adds a little breathing room between the two tables

    % --- TABLE B: PER PROTOCOL SET ---
    \begin{tabularx}{\textwidth}{@{} X c c l @{}} 
        \toprule
        \textbf{B. Per Protocol Set (PPS)} & \textbf{Internet} & \textbf{LLM} & \textbf{Balance Test} \\ 
        \midrule

        % --- BIO-EXP TOTAL ---
        \textbf{BIO-EXP total (0--30)} & & & \\
        \hspace{3mm} Mean (SD) & 4.3 (5.7) & 4.6 (4.2) & $t(124.41) = -0.32, P = 0.752, d = -0.056$ \\
        \hspace{3mm} Median (IQR) & 3.0 (7.0) & 3.5 (6.0) & \\
        \addlinespace

        % --- PIPETTING ---
        \textbf{Pipetting subscore (0--6)} & & & \\
        \hspace{3mm} Mean (SD) & 2.3 (2.2) & 2.4 (2.0) & $t(125.44) = -0.47, P = 0.642, d = -0.082$ \\
        \hspace{3mm} Median (IQR) & 2.0 (4.0) & 2.0 (4.0) & \\
        \addlinespace

        % --- CELL CULTURE ---
        \textbf{Cell culture subscore (0--6)} & & & \\
        \hspace{3mm} Mean (SD) & 0.4 (0.8) & 0.4 (0.7) & $t(123.28) = -0.25, P = 0.806, d = -0.043$ \\
        \hspace{3mm} Median (IQR) & 0.0 (0.2) & 0.0 (1.0) & \\
        \addlinespace

        % --- MOLECULAR CLONING ---
        \textbf{Molecular cloning subscore (0--6)} & & & \\
        \hspace{3mm} Mean (SD) & 0.6 (1.3) & 0.6 (1.3) & $t(125.44) = -0.07, P = 0.946, d = -0.012$ \\
        \hspace{3mm} Median (IQR) & 0.0 (0.2) & 0.0 (1.0) & \\
        \addlinespace

        % --- TRANSFECTION ---
        \textbf{Transfection / AAV subscore (0--6)} & & & \\
        \hspace{3mm} Mean (SD) & 0.2 (0.6) & 0.1 (0.3) & $t(93.44) = 0.96, P = 0.338, d = 0.170$ \\
        \hspace{3mm} Median (IQR) & 0.0 (0.0) & 0.0 (0.0) & \\
        \addlinespace

        % --- qPCR ---
        \textbf{qPCR subscore (0--6)} & & & \\
        \hspace{3mm} Mean (SD) & 0.9 (1.1) & 1.0 (1.2) & $t(125.57) = -0.54, P = 0.591, d = -0.095$ \\
        \hspace{3mm} Median (IQR) & 1.0 (1.0) & 1.0 (2.0) & \\
        \bottomrule
    \end{tabularx}
\end{table}

\emph{Note: BIO-EXP: Sum of 5 task subscores (each 0 -- 6); total 0 -- 30. Higher equals more prior biology experience. Balance tests were conducted rowwise using Welch's t-tests. Effect size was quantified using Cohen's \(d\).}

%% file: tables/ed-table-7.tex
\begin{table}[H]
    \caption{Self-Reported Measures}
    \label{tab:ed-table-7}
    \centering

    % --- TABLE A: FULL ANALYSIS SET ---
    \begin{tabularx}{\textwidth}{@{} X c c l @{}} 
        \toprule
        \textbf{A. Full Analysis Set (FAS)} & \textbf{Internet} & \textbf{LLM} & \textbf{Balance Test} \\ 
        \midrule

        % --- TAM SECTION ---
        \textbf{LLM Tech Acceptance (TAM)} & & & \\
        \textit{TAM overall} & & & $t(145.98) = 0.46, P = 0.648, d = 0.075$ \\
        \hspace{3mm} Mean (SD) & 3.9 (0.6) & 3.8 (0.7) & \\
        \hspace{3mm} Median (IQR) & 3.9 (0.9) & 3.9 (0.8) & \\
        \addlinespace

        \textit{Perceived Usefulness} & & & $t(145.01) = 0.80, P = 0.424, d = 0.132$ \\
        \hspace{3mm} Mean (SD) & 4.0 (0.8) & 3.8 (0.9) & \\
        \hspace{3mm} Median (IQR) & 4.0 (1.3) & 3.8 (1.3) & \\
        \addlinespace

        \textit{Perceived Ease of Use} & & & $t(145.82) = -0.14, P = 0.890, d = -0.023$ \\
        \hspace{3mm} Mean (SD) & 3.8 (0.7) & 3.8 (0.7) & \\
        \hspace{3mm} Median (IQR) & 3.8 (0.8) & 3.8 (0.8) & \\
        \addlinespace

        % --- NGSE SECTION ---
        \textbf{New Generalized Self-Efficacy (NGSE)} & & & \\
        \textit{NGSE score} & & & $t(146.00) = -0.40, P = 0.691, d = -0.066$ \\
        \hspace{3mm} Mean (SD) & 3.8 (0.6) & 3.9 (0.6) & \\
        \hspace{3mm} Median (IQR) & 3.9 (0.8) & 3.9 (0.9) & \\
        \addlinespace

        % --- PERFECTIONISM SECTION ---
        \textbf{Perfectionism (FMPS-B)} & & & \\
        \textit{FMPS-B total (8--40)} & & & $t(145.64) = 1.03, P = 0.305, d = 0.169$ \\
        \hspace{3mm} Mean (SD) & 24.7 (5.2) & 23.9 (5.1) & \\
        \hspace{3mm} Median (IQR) & 24.0 (6.0) & 23.0 (7.0) & \\
        \addlinespace

        \textit{Evaluative Concerns} & & & $t(146.00) = 1.08, P = 0.284, d = 0.177$ \\
        \hspace{3mm} Mean (SD) & 11.0 (3.7) & 10.3 (3.8) & \\
        \hspace{3mm} Median (IQR) & 11.0 (6.0) & 10.0 (4.5) & \\
        \addlinespace

        \textit{Strivings} & & & $t(145.94) = 0.38, P = 0.705, d = 0.062$ \\
        \hspace{3mm} Mean (SD) & 13.7 (3.3) & 13.5 (3.3) & \\
        \hspace{3mm} Median (IQR) & 14.0 (4.0) & 14.0 (4.0) & \\
        \bottomrule
    \end{tabularx}
\end{table}

\begin{table}[H]
    % --- TABLE B: PER PROTOCOL SET ---
    \begin{tabularx}{\textwidth}{@{} X c c l @{}} 
        \toprule
        \textbf{B. Per Protocol Set (PPS)} & \textbf{Internet} & \textbf{LLM} & \textbf{Balance Test} \\ 
        \midrule

        % --- TAM SECTION ---
        \textbf{LLM Tech Acceptance (TAM)} & & & \\
        \textit{TAM overall} & & & $t(125.68) = 0.15, P = 0.879, d = 0.027$ \\
        \hspace{3mm} Mean (SD) & 3.9 (0.7) & 3.8 (0.7) & \\
        \hspace{3mm} Median (IQR) & 3.9 (1.0) & 3.9 (0.9) & \\
        \addlinespace

        \textit{Perceived Usefulness} & & & $t(124.30) = 0.51, P = 0.610, d = 0.090$ \\
        \hspace{3mm} Mean (SD) & 4.0 (0.8) & 3.9 (0.9) & \\
        \hspace{3mm} Median (IQR) & 4.0 (1.2) & 4.0 (1.6) & \\
        \addlinespace

        \textit{Perceived Ease of Use} & & & $t(126.00) = -0.36, P = 0.723, d = -0.063$ \\
        \hspace{3mm} Mean (SD) & 3.8 (0.7) & 3.8 (0.7) & \\
        \hspace{3mm} Median (IQR) & 3.8 (1.0) & 3.8 (0.8) & \\
        \addlinespace

        % --- NGSE SECTION ---
        \textbf{New Generalized Self-Efficacy (NGSE)} & & & \\
        \textit{NGSE score} & & & $t(125.52) = -0.75, P = 0.455, d = -0.133$ \\
        \hspace{3mm} Mean (SD) & 3.8 (0.6) & 3.9 (0.6) & \\
        \hspace{3mm} Median (IQR) & 3.9 (0.8) & 3.9 (0.9) & \\
        \addlinespace

        % --- PERFECTIONISM SECTION ---
        \textbf{Perfectionism (FMPS-B)} & & & \\
        \textit{FMPS-B total} & & & $t(125.66) = 0.90, P = 0.368, d = 0.160$ \\
        \hspace{3mm} Mean (SD) & 24.9 (5.5) & 24.0 (5.2) & \\
        \hspace{3mm} Median (IQR) & 24.0 (6.5) & 23.5 (7.0) & \\
        \addlinespace

        \textit{Evaluative Concerns} & & & $t(125.99) = 0.96, P = 0.336, d = 0.171$ \\
        \hspace{3mm} Mean (SD) & 11.2 (3.8) & 10.6 (3.9) & \\
        \hspace{3mm} Median (IQR) & 11.0 (5.0) & 10.0 (5.0) & \\
        \addlinespace

        \textit{Strivings} & & & $t(126.00) = 0.34, P = 0.738, d = 0.059$ \\
        \hspace{3mm} Mean (SD) & 13.6 (3.4) & 13.4 (3.4) & \\
        \hspace{3mm} Median (IQR) & 13.5 (4.2) & 13.5 (5.0) & \\
        \bottomrule
    \end{tabularx}
\end{table}

\emph{Note:  TAM overall: Mean of 12 items (1 -- 5). PU and PEOU are 6 item means (1 -- 5). Higher equals more acceptance. NGSE: Mean of 8 items (1 -- 5). Higher equals greater self-efficacy. FMPS B: Total = EC + Strivings (8 -- 40). Higher equals more perfectionism. Balance tests were conducted rowwise using Welch's t-tests. Effect size was quantified using Cohen’s \(d\).}

%% file: tables/ed-table-8.tex
\begin{table}[H]
    \caption{Primary and Secondary Outcomes}
    \label{tab:ed-table-8}
    \centering

    % Reduce padding between columns slightly (default is usually 6pt) to give text more room
    \setlength{\tabcolsep}{3pt} 

    \begin{tabularx}{\textwidth}{@{} X l l l @{}} 
        \toprule
        % Top Header: Left the last column empty to save width/clutter
        & \multicolumn{2}{c}{\textbf{Effect Estimates (LLM Intervention)}} & \\ 
        \cmidrule(lr){2-3} 
        \textit{Outcomes} & \textbf{Risk Diff. (95\% CI)} & \textbf{Risk Ratio (95\% CI)} & \textbf{Fisher's Exact} $P$ \\ 
        \midrule

        % --- SECTION A: FAS ---
        \multicolumn{4}{@{}l}{\textbf{A. Full Analysis Set (FAS)}} \\ 
        \addlinespace[0.5em]

        \textbf{Primary Outcome} \newline (Tasks 2--4) & -1.4\% (-9.9\%, 6.9\%) & 0.790 (0.237, 2.624) & 0.759 \\ 
        \addlinespace

        Pre-task 1. Micropipetting & 4.2\% (-8.6\%, 16.9\%) & 1.054 (0.894, 1.249) & 0.329 \\ 
        Task 2. Cell Culture & 13.6\% (-1.8\%, 28.1\%) & 1.246 (0.971, 1.619) & 0.059 \\ 
        Task 3. Molecular Cloning & -2.8\% (-14.9, 9.2\%) & 0.846 (0.423, 1.685) & 0.752 \\ 
        Task 4. Virus Production & 6.3\% (-5.2\%, 17.8\%) & 1.535 (0.724, 3.293) & 0.192 \\ 
        Task 5. RNA Quantification & 3.8\% (-5.6\%, 13.4\%) & 1.579 (0.569, 4.427) & 0.290 \\ 

        \midrule 
        % --- SECTION B: PPS ---
        \multicolumn{4}{@{}l}{\textbf{B. Per Protocol Set (PPS)}} \\ 
        \addlinespace[0.5em]

        \textbf{Primary Outcome} \newline (Tasks 2--4) & -1.56\% (-11.52\%, 8.24\%) & 0.80 (0.241, 2.640) & 0.754 \\ 
        \addlinespace

        Pre-task 1. Micropipetting & 3.13\% (-10.1\%, 16.31\%) & 1.038 (0.880, 1.232) & 0.408 \\ 
        Task 2. Cell Culture & 17.19\% (1.46\%, 31.84\%) & 1.275 (1.022, 1.625) & 0.025 \\ 
        Task 3. Molecular Cloning & -3.13\% (-17.0\%, 10.87\%) & 0.857 (0.434, 1.584) & 0.745 \\ 
        Task 4. Virus Production & 6.25\% (-7.02\%, 19.37\%) & 1.444 (0.679, 3.102) & 0.241 \\ 
        Task 5. RNA Quantification & 4.69\% (-6.32\%, 15.89\%) & 1.600 (0.581, 4.454) & 0.280 \\ 

        \bottomrule
    \end{tabularx}
\end{table}

\emph{Note: Confidence Intervals: For risk difference, Newcombe's method (Wilson-Wilson) is used, for risk proportion (relative risk) we use the Koopman exact interval, and for odds ratio we use Cornfield's method.}

%% file: tables/ed-table-9.tex
\begin{table}[H]
    \caption{LOO Summary}
    \label{tab:ed-table-9}
    \centering

    % Define columns: 
    % 1. X (Left aligned, fills 1/3 space)
    % 2. >{\centering\arraybackslash}X (Centered, fills 1/3 space)
    % 3. >{\centering\arraybackslash}X (Centered, fills 1/3 space)
    \begin{tabularx}{\textwidth}{@{} X >{\centering\arraybackslash}X >{\centering\arraybackslash}X @{}}
        \toprule
        & \multicolumn{2}{c}{\textbf{Estimate ($\bm{\pm}$SE)}} \\
        \cmidrule(lr){2-3} 
        \textbf{Model} & \textbf{elpd\_loo} & \textbf{p\_loo} \\ 
        \midrule

        % --- SECTION A: FAS ---
        \multicolumn{3}{@{}l}{\textbf{A. Full Analysis Set (FAS)}} \\ 
        \addlinespace[0.5em]

        Individual: hierarchical & -309.0 ($\pm$15.6) & 94.4 ($\pm$5.5) \\ 
        Individual: independent & -310.8 ($\pm$15.9) & 97.9 ($\pm$5.7) \\ 
        Task: hierarchical & -362.7 ($\pm$15.9) & 7.2 ($\pm$0.5) \\ 
        Task: independent & -364.9 ($\pm$16.3) & 10.0 ($\pm$0.7) \\ 

        \midrule 

        % --- SECTION B: PPS ---
        \multicolumn{3}{@{}l}{\textbf{B. Per Protocol Set (PPS)}} \\ 
        \addlinespace[0.5em]

        Individual: hierarchical & -269.7 ($\pm$14.5) & 75.7 ($\pm$4.8) \\ 
        Individual: independent & -271.1 ($\pm$15.0) & 80.3 ($\pm$5.3) \\ 
        Task: hierarchical & -305.9 ($\pm$14.7) & 7.4 ($\pm$0.5) \\ 
        Task: independent & -307.6 ($\pm$15.2) & 9.9 ($\pm$0.7) \\ 

        \bottomrule
    \end{tabularx}
\end{table}

\emph{Estimate ± SD by model and dataset. ELPD LOO: Bayesian leave-one-out cross-validation estimate of the expected log pointwise predictive density, a measure of the model's generalization to out-of-sample data. P LOO: Effective number of parameters (estimated via leave-one-out cross-validation)}

%% file: tables/ed-table-10.tex
\begin{table}[H]
    \caption{RMST Analysis for Time-to-Success}
    \label{tab:ed-table-10a}
    \centering
    \setlength{\tabcolsep}{4pt} % Slightly tighter spacing to fit CIs comfortably

    \begin{tabularx}{\textwidth}{@{} X c l l l c @{}}
        \toprule
        & & \multicolumn{2}{c}{\textbf{RMST (95\% CI)}} & \multicolumn{2}{c}{\textbf{Comparisons}} \\
        \cmidrule(lr){3-4} \cmidrule(l){5-6}
        \textbf{Task} & $\bm{\tau}$ & \textbf{INT} & \textbf{LLM} & \textbf{Diff. (95\% CI)} & $\bm{P}$ \\ 
        \midrule

        \textbf{Pre-Task 1} \newline Micropipetting & 53.17 & 14.53 \newline (9.54, 19.53) & 11.28 \newline (6.49, 16.07) & -3.25 \newline (-10.17, 3.67) & 0.36 \\ 
        \addlinespace

        \textbf{Task 2} \newline Cell Culture & 53.18 & 39.30 \newline (35.82, 42.78) & 33.29 \newline (29.65, 36.92) & -6.02 \newline (-11.05, -0.98) & 0.02* \\ 
        \addlinespace

        \textbf{Task 3} \newline Molecular Cloning & 44.38 & 41.80 \newline (40.43, 43.16) & 42.21 \newline (40.86, 43.56) & 0.41 \newline (-1.51, 2.33) & 0.68 \\ 
        \addlinespace

        \textbf{Task 4} \newline Virus Production & 39.38 & 37.02 \newline (35.38, 38.67) & 35.61 \newline (33.59, 37.63) & -1.41 \newline (-4.01, 1.19) & 0.29 \\ 
        \addlinespace

        \textbf{Task 5} \newline RNA Quantification & 23.38 & 22.86 \newline (22.32, 23.40) & 22.09 \newline (21.12, 23.06) & -0.77 \newline (-1.88, 0.34) & 0.18 \\ 

        \textbf{Primary Outcome} \newline (Tasks 2--4) & 53.01 & 51.93 \newline (50.92, 52.94) & 52.04 \newline (51.05, 53.03) & 0.11 \newline (-1.30, 1.52) & 0.88 \\

        \bottomrule
    \end{tabularx}
\end{table}

\begin{table}[H]
    \caption{Median Time \& Log-Rank Analysis}
    \label{tab:ed-table-10b}
    \centering
    \setlength{\tabcolsep}{6pt} % More breathing room for this simpler table

    \begin{tabularx}{\textwidth}{@{} X l l l @{}}
        \toprule
        & \multicolumn{2}{c}{\textbf{Median Time (95\% CI)}} & \textbf{Log-Rank Test} \\
        \cmidrule(lr){2-3} 
        \textbf{Task} & \textbf{INT} & \textbf{LLM} & $\bm{P}$ \textbf{(df=1, 1-sided)} \\ 
        \midrule

        \textbf{Pre-Task 1} \newline Micropipetting & 3.07 (2.14, 7.06) & 2.04 (1.14, 2.11) & 0.016* \\ 
        \addlinespace

        \textbf{Task 2} \newline Cell Culture & 45.19 (35.11, N/A) & 31.12 (21.11, 42.18) & 0.012* \\ 
        \addlinespace

        \textbf{Task 3} \newline Molecular Cloning & N/A & N/A & 0.32 \\ 
        \addlinespace

        \textbf{Task 4} \newline Virus Production & N/A & N/A & 0.13 \\ 
        \addlinespace

        \textbf{Task 5} \newline RNA Quantification & N/A & N/A & 0.18 \\ 

        \textbf{Primary Outcome} \newline (Tasks 2--4) & N/A & N/A & N/A \\

        \bottomrule
    \end{tabularx}
\end{table}

%% file: tables/ed-table-11.tex
\begin{table}[H]
    \caption{Median Number of Attempts to Succeed per Task}
    \label{tab:ed-table-11}
    \centering

    % 5 columns: 
    % 1. X (Task Name)
    % 2-3. c (FAS Data)
    % 4-5. c (PPS Data)
    \begin{tabularx}{\textwidth}{@{} X c c c c @{}}
        \toprule
        & \multicolumn{2}{c}{\textbf{Full Analysis Set (FAS)}} & \multicolumn{2}{c}{\textbf{Per Protocol Set (PPS)}} \\
        \cmidrule(lr){2-3} \cmidrule(l){4-5}
        \textbf{Task} & \textbf{Internet} & \textbf{LLM} & \textbf{Internet} & \textbf{LLM} \\ 
        \midrule

        Pre-task 1. Micropipetting & 8 [9] & 5 [4] & 8 [8] & 4.5 [4] \\ 
        Task 2. Cell Culture       & 4 [6] & 3 [3] & 4 [4] & 3 [3] \\ 
        Task 3. Molecular Cloning  & 2 [2] & 2 [2] & 2 [2] & 2 [1.5] \\ 
        Task 4. Virus Production   & 2 [3] & 2 [2] & 2 [2] & 2 [2] \\ 
        Task 5. RNA Quantification & 2 [1] & 2 [2] & 2 [0] & 2 [1.5] \\ 

        \bottomrule
    \end{tabularx}
\end{table}

%% file: tables/ed-table-12.tex
\begin{table}[H]
\caption{Per-task Milestone Achievement Rates}
\label{tab:ed-table-12}
\centering
\small % 'small' is standard for tables; larger than 'scriptsize'
\setlength{\tabcolsep}{3pt} % Reduces whitespace to allow text to expand

% Column Layout: X (Task Name) + 5 data columns (l)
% The X column will naturally take less space now that the text in 'l' columns is larger
\begin{tabularx}{\textwidth}{@{} X l l l l l @{}}
\toprule
& \textbf{INT n/N} & \textbf{LLM n/N} & \textbf{Risk Diff.} & \textbf{Risk Ratio} & \textbf{Fisher's} \\
\textbf{Task} & \textbf{(95\% CI)} & \textbf{(95\% CI)} & \textbf{(95\% CI)} & \textbf{(95\% CI)} & \textbf{Exact} $\bm{P}$ \\ 
\midrule

% --- SECTION A: FAS ---
\multicolumn{6}{@{}l}{\textbf{A. Full Analysis Set (FAS)}} \\ 
\addlinespace[0.5em]

Task 2 & 64.5\% \newline (53.3, 74.3\%) & 75.3\% \newline (64.6, 83.6\%) & 10.85\% \newline (-0.04, 0.25\%) & 1.17 \newline (0.95, 1.46) & 0.099 \\ 
\addlinespace

Task 3 & 23.7\% \newline (15.5, 34.4\%) & 28.6\% \newline (19.7, 39.5\%) & 4.89\% \newline (-0.09, 0.19\%) & 1.21 \newline (0.71, 2.06) & 0.307 \\ 
\addlinespace

Task 4 & 13.2\% \newline (7.3, 22.6\%) & 13.0\% \newline (7.2, 22.3\%) & -0.17\% \newline (-0.11, 0.11\%) & 0.99 \newline (0.44, 2.19) & 0.607 \\ 
\addlinespace

Task 5 & 9.2\% \newline (4.5, 17.8\%) & 18.2\% \newline (11.2, 28.2\%) & 8.97\% \newline (-0.02, 0.20\%) & 1.97 \newline (0.87, 4.55) & 0.084 \\ 

\midrule 

% --- SECTION B: PPS ---
\multicolumn{6}{@{}l}{\textbf{B. Per Protocol Set (PPS)}} \\ 
\addlinespace[0.5em]

Task 2 & 73.4\% \newline (61.5, 82.7\%) & 85.9\% \newline (75.4, 92.4\%) & 12.50\% \newline (-1.54, 26.07\%) & 1.17 \newline (0.98, 1.42) & 0.062 \\
\addlinespace

Task 3 & 26.6\% \newline (17.3, 38.5\%) & 34.4\% \newline (23.9, 46.6\%) & 7.81\% \newline (-8.04, 23.15\%) & 1.29 \newline (0.77, 2.20) & 0.221 \\
\addlinespace

Task 4 & 15.6\% \newline (8.7, 26.4\%) & 15.6\% \newline (8.7, 26.4\%) & 0.00\% \newline (-12.82, 12.82\%) & 1.00 \newline (0.46, 2.20) & 0.596 \\
\addlinespace

Task 5 & 10.9\% \newline (5.4, 20.9\%) & 21.9\% \newline (13.5, 33.4\%) & 10.94\% \newline (-2.07, 23.75\%) & 2.00 \newline (0.89, 4.57) & 0.075 \\

\bottomrule
\end{tabularx}
\end{table}

\emph{Note: Task 2: Cell Culture; Task 3: Molecular Cloning; Task 4: Virus Production; Task 5: RNA Quantification.}

%% file: tables/ed-table-13.tex
\begin{table}[h]
\caption{Non-Verbal Reasoning}
\label{tab:ed-table-13}
\centering

% 4 columns:
% 1. X (Description, left aligned)
% 2. c (Internet Data)
% 3. c (LLM Data)
% 4. c (Total N)
\begin{tabularx}{\textwidth}{@{} X c c c @{}}
\toprule
& \textbf{Internet} & \textbf{LLM} & \textbf{Total N*} \\ 
\midrule
Raven's 2 Standard score --- mean (SD) & 115.5 (12.8) & 114.4 (12.2) & 127 \\ 
\bottomrule
\end{tabularx}
\end{table}

Raven's 2 Digital Short Form: M=100, SD=15

\emph{Note: *Total refers to the number of participants who took the Raven's 2 Digital Short Form Progressive Matrices Test. This test was administered at the end of the study so participants do not fall directly into FAS or PPS.}

%% file: tables/ed-table-14.tex
\begin{table}[h]
\caption{Subgroup Analyses}
\label{tab:ed-table-14}
\centering
\small % Reduces font size slightly to fit the columns
\begin{tabularx}{\textwidth}{@{} >{\raggedright\arraybackslash}X >{\raggedright\arraybackslash}X >{\raggedright\arraybackslash}X c >{\raggedright\arraybackslash}X >{\raggedright\arraybackslash}X @{}}
\toprule
\textbf{Covariate (Full Analysis Set Data)} & 
\textbf{Main Effect OR (95\% CI), \textit{P}} & 
\textbf{Interaction ROR* (95\% CI), \textit{P}} & 
\textbf{LRT**} & 
\textbf{Stratified Effect: Internet, OR (95\% CI), \textit{P}} & 
\textbf{Stratified Effect: LLM} \\ % Note: Metric implied from context
\midrule

Prior Biology Experience & 
1.02 (0.72--1.43), \textit{P} = 0.930 & 
0.68 (0.34--1.36), \textit{P} = 0.274 & 
\(P = 0.272, \chi^2 = 1.21\) & 
1.21 (0.75--1.95), \textit{P} = 0.429 & 
0.82 (0.50--1.36), \textit{P} = 0.448 \\
\addlinespace

Self-Reported Prior LLM Experience & 
1.05 (0.75--1.48), \textit{P} = 0.777 & 
0.57 (0.28--1.16), \textit{P} = 0.123 & 
\(P = 0.119, \chi^2 = 2.43\) & 
1.35 (0.84--2.17), \textit{P} = 0.213 & 
0.78 (0.46--1.31), \textit{P} = 0.342 \\
\addlinespace

Non-Verbal Reasoning & 
1.91 (1.24--2.94), \textit{P} = 0.003** & 
0.94 (0.39--2.27), \textit{P} = 0.898 & 
\(P = 0.898, \chi^2 = 0.02\) & 
1.95 (1.11--3.44), \textit{P} = 0.020* & 
1.84 (0.94--3.62), \textit{P} = 0.075 \\
\addlinespace

Number of Tokens (LLM Only) & 
1.39 (0.71--2.73), \textit{P} = 0.335 & 
n/a & 
n/a & 
n/a & 
n/a \\
\addlinespace

Number of Uploaded Images (LLM Only) & 
0.95 (0.59--1.52), \textit{P} = 0.818 & 
n/a & 
n/a & 
n/a & 
n/a \\
\addlinespace

YouTube Searches & 
1.45 (0.96--2.18), \textit{P} = 0.074 & 
0.66 (0.27--1.65), \textit{P} = 0.376 & 
\(P = 0.380, \chi^2 = 0.77\) & 
1.62 (0.99--2.67), \textit{P} = 0.057 & 
1.08 (0.52--2.30), \textit{P} = 0.849 \\

\bottomrule
\end{tabularx}
\end{table}

\emph{Note: Effect sizes reported as odds ratios per SD (with 95\% CI). *ROR: Ratio of odds ratios, a measure of how the odds ratio changes as a function of the moderator. **LRT: Global likelihood ratio test comparing model with arm effects against models without; an insignificant test indicates no support for moderation. Stratified effects refer to the effect of the covariate within an arm stratum.}

%% file: tables/ed-table-15.tex
\begin{table}[h]
\caption{Procedural Steps for Tasks 2--5}
\label{tab:ed-table-15}
\centering
\small % Slightly smaller text handles the descriptive lists better
\begin{tabularx}{\textwidth}{@{} l X @{}}
\toprule
\textbf{Step} & \textbf{Description of Milestone / Success Criteria} \\ 
\midrule

% --- TASK 2 ---
\multicolumn{2}{@{}l}{\textbf{Task 2. Cell Culture}} \\ 
\addlinespace[0.2em]
0 & Never thawed \\
1 & Self-reported task start \\
2 & Thawed, but only got score 0 / invalid (no successful P1) \\
3 & Successful P1 ($\ge \qty{70}{\percent}$ viability), but not P2 (milestone) \\
4 & Successful P2 ($\ge \qty{85}{\percent}$ viability), but not P3 \\
5 & Successful P3 ($\ge \qty{85}{\percent}$ viability) (task success) \\

\midrule
% --- TASK 3 ---
\multicolumn{2}{@{}l}{\textbf{Task 3. Molecular Cloning}} \\ 
\addlinespace[0.2em]
0 & No request/attempt \\
1 & Self-reported task start \\
2 & Requested materials (not necessarily correct materials) \\
3 & Requested the correct materials \\
4 & Generated plasmid \& submitted for sequencing \\
5 & Analyzed sequence and submitted for outcome assessment \\
6 & At least one correct plasmid (0--10 mutations) (milestone) \\
7 & Two correct plasmids with 0 mutations (task success) \\

\midrule
% --- TASK 4 ---
\multicolumn{2}{@{}l}{\textbf{Task 4. Virus Production}} \\ 
\addlinespace[0.2em]
0 & No request/attempt \\
1 & Self-reported task start \\
2 & Requested materials, but not necessarily correct \\
3 & Requested the correct materials \\
4 & $> \qty{5}{\percent}$ GFP transfection efficiency \\
5 & $\qty{2e10}{gc/mL}$ (task success) \\

\midrule
% --- TASK 5 ---
\multicolumn{2}{@{}l}{\textbf{Task 5. RNA Quantification}} \\ 
\addlinespace[0.2em]
0 & No request/attempt \\
1 & Self-reported task start \\
2 & Requested materials, but not necessarily correct \\
3 & Requested the correct materials \\
4 & Correct standard curve (milestone) \\
5 & Standard curve plus correct unknown concentration (task success) \\

\bottomrule
\end{tabularx}
\end{table}

%% file: tables/ed-table-16.tex
\begin{table}[H]
\caption{NASA-TLX}
\label{tab:ed-table-16}
\centering
\small 
\setlength{\tabcolsep}{4pt}

% Layout: 
% 1. X (Label column)
% 2-5. >{\centering\arraybackslash}X (Data columns, centered and equal width)
\begin{tabularx}{\textwidth}{@{} X >{\centering\arraybackslash}X >{\centering\arraybackslash}X >{\centering\arraybackslash}X >{\centering\arraybackslash}X @{}}
\toprule
& \multicolumn{2}{c}{\textbf{Internet}} & \multicolumn{2}{c}{\textbf{LLM}} \\
\cmidrule(lr){2-3} \cmidrule(l){4-5}
& \textbf{Week 1} & \textbf{Week 8} & \textbf{Week 1} & \textbf{Week 8} \\ 
\midrule

% --- SECTION A: FAS ---
\multicolumn{5}{@{}l}{\textbf{A. Full Analysis Set (FAS)}} \\ 
\addlinespace[0.2em]

\textbf{Mean (SD)} & & & & \\
\hspace{3mm} Mental & 5.2 (1.8) & 5.2 (1.9) & 5.3 (1.6) & 5.0 (2.3) \\ 
\hspace{3mm} Physical & 3.4 (1.9) & 3.8 (2.0) & 3.3 (1.7) & 3.3 (2.0) \\ 
\hspace{3mm} Temporal & 3.7 (1.6) & 4.1 (1.9) & 3.6 (1.6) & 4.0 (2.2) \\ 
\hspace{3mm} Performance & 5.0 (1.5) & 4.5 (1.8) & 5.8 (1.5) & 5.1 (1.9) \\ 
\hspace{3mm} Effort & 5.7 (1.7) & 5.4 (2.0) & 5.7 (1.7) & 5.1 (2.0) \\ 
\hspace{3mm} Frustration & 5.1 (2.0) & 4.5 (2.0) & 5.1 (1.8) & 4.1 (2.2) \\ 
\addlinespace[0.5em]

\textbf{Median (IQR)} & & & & \\
\hspace{3mm} Mental & 5.0 (3.5) & 5.0 (3.0) & 5.0 (2.1) & 5.0 (4.9) \\ 
\hspace{3mm} Physical & 3.0 (3.0) & 3.5 (3.0) & 3.0 (2.2) & 2.0 (4.0) \\ 
\hspace{3mm} Temporal & 3.0 (3.0) & 4.0 (3.0) & 3.0 (2.5) & 3.0 (4.0) \\ 
\hspace{3mm} Performance & 5.0 (2.0) & 5.0 (3.0) & 5.5 (2.1) & 5.0 (3.0) \\ 
\hspace{3mm} Effort & 5.5 (2.0) & 5.0 (3.0) & 6.0 (3.0) & 5.0 (3.9) \\ 
\hspace{3mm} Frustration & 4.8 (3.9) & 4.0 (3.0) & 5.0 (2.6) & 3.8 (4.0) \\ 

\midrule 

% --- SECTION B: PPS ---
\multicolumn{5}{@{}l}{\textbf{B. Per Protocol Set (PPS)}} \\ 
\addlinespace[0.2em]

\textbf{Mean (SD)} & & & & \\
\hspace{3mm} Mental & 5.1 (1.6) & 5.2 (1.9) & 5.2 (1.5) & 4.9 (2.2) \\ 
\hspace{3mm} Physical & 3.3 (1.9) & 3.8 (2.1) & 3.2 (1.6) & 3.2 (1.9) \\ 
\hspace{3mm} Temporal & 3.7 (1.6) & 4.1 (1.9) & 3.6 (1.5) & 3.9 (2.2) \\ 
\hspace{3mm} Performance & 5.1 (1.5) & 4.4 (1.8) & 5.8 (1.5) & 5.1 (1.9) \\ 
\hspace{3mm} Effort & 5.6 (1.7) & 5.4 (1.9) & 5.7 (1.8) & 5.0 (2.0) \\ 
\hspace{3mm} Frustration & 5.0 (1.9) & 4.5 (2.0) & 5.0 (1.7) & 4.2 (2.2) \\ 
\addlinespace[0.5em]

\textbf{Median (IQR)} & & & & \\
\hspace{3mm} Mental & 5.0 (3.6) & 5.0 (3.0) & 5.0 (2.5) & 5.0 (4.9) \\ 
\hspace{3mm} Physical & 2.8 (2.2) & 3.0 (3.5) & 2.8 (2.0) & 2.0 (3.0) \\ 
\hspace{3mm} Temporal & 3.0 (3.0) & 3.5 (3.0) & 3.0 (2.1) & 3.0 (4.0) \\ 
\hspace{3mm} Performance & 5.0 (2.2) & 4.0 (3.0) & 6.0 (2.0) & 5.0 (3.0) \\ 
\hspace{3mm} Effort & 5.5 (2.5) & 5.0 (3.0) & 5.8 (3.1) & 5.0 (3.8) \\ 
\hspace{3mm} Frustration & 4.5 (3.6) & 4.0 (3.0) & 4.8 (3.1) & 3.8 (4.0) \\ 

\bottomrule
\end{tabularx}
\end{table}

\emph{Note: NASA-TLX items were administered daily at the task level for each task a participant reported working on. For each participant, all NASA-TLX measurements within a given week were averaged to produce a single weekly score. Weekly scores were then aggregated across participants within each treatment arm. Weeks exclude enrollment/training, and only include time spent working on tasks.}

%% file: tables/ed-table-17.tex
\begin{table}[H]
\caption{Plasmid Assembly Variant Classes}
\label{tab:ed-table-17}
\centering
\begin{tabularx}{\textwidth}{@{} X @{}}
\toprule
\textbf{Description of Variant Classes} \\ 
\midrule

The following variant classes are not considered mismatches for the purpose of Task 3 (Molecular Cloning) quality criterion: \\
\addlinespace[0.5em]

\begin{itemize}
    \setlength\itemsep{0.5em} % Adds a little breathing room between bullets
    \item \textbf{Nanopore sequencing artifacts:}
    \begin{itemize}
        \item Deletions in homopolymer stretches, defined as a stretch of identical nucleotides greater than or equal to 8 base pairs.
        \item Errors at the Dam methylation site GATC.
        \item Errors at the middle position of the Dcm methylation site CCTGG or CCAGG.
    \end{itemize}
    
    \item \textbf{Silent mutations:} Synonymous mutations that do not alter the protein coding sequence.
    
    \item \textbf{Mutations in bacterial backbone regions:} (e.g., origins of replication or antibiotic resistance) defined as the portion of the plasmid outside the mammalian promoters, terminators, coding regions, and viral genome.
\end{itemize} \\

\bottomrule
\end{tabularx}
\end{table}

%% file: tables/ed-table-18.tex
\begin{table}[H]
\caption{Self-Reported Protocol Adherence on LLM Usage}
\label{tab:ed-table-18}
\centering
\setlength{\tabcolsep}{6pt} 

% Layout: X (Response Category) + l (Count)
\begin{tabularx}{\textwidth}{@{} X l @{}}
\toprule

% 1. Question text is now the very first thing (spanning both columns)
\multicolumn{2}{@{}p{\textwidth}@{}}{\textbf{Question:} During the study, did you ever use an AI or large-language-model (LLM) tool (e.g., ChatGPT, Claude, Gemini, Copilot)?} \\ 
\addlinespace[0.8em] % Add distinct separation before the sub-headers

% 2. Column Headers are now below the question
\textbf{Response} & \textbf{Internet Arm (n = 76)} \\ 
\midrule

% 3. Data
No Data & 7 \\ 
No, Never & 65 \\ 
Yes, once & 2 \\ 
Yes, occasionally ($\le$25\% of sessions) & 2 \\ 
Yes, often ($>$25\% of sessions) & 0 \\ 

\bottomrule
\end{tabularx}
\end{table}

\emph{Note: The Participant Adherence Survey was administered at either the end-of-study, or upon discontinuation and/or withdrawal. No data represents participants who declined to complete the survey upon withdrawal of consent, or those who were lost to follow-up.}